\documentclass[amsmath,amssymb,aps,superscriptaddress,preprint]{revtex4-1}
\usepackage{graphicx}
\usepackage{epstopdf}

\usepackage[usenames, dvipsnames]{color}

\usepackage[normalem]{ulem}

	\newcommand{\QQ}{\mathbf{Q}}
	\newcommand{\KK}{\mathbf{K}}
	\newcommand{\kk}{\mathbf{k}}
	\newcommand{\qq}{\mathbf{q}}
	\newcommand{\GG}{\mathbf{G}}
	\newcommand{\bb}{\mathbf{b}}
	\newcommand{\rr}{\mathbf{r}}
	\newcommand{\kkappa}{\boldsymbol{\kappa}}
	\newcommand{\rrho}{\boldsymbol{\rho}}
	\newcommand{\ket}[1]{| #1 \rangle}

	\renewcommand{\exp}[1]{{\rm e}^{#1}}
	\newcommand{\Ams}{\mathrm{\AA}}

	\newcommand{\ud}{\mathrm{d}}
	\renewcommand{\exp}[1]{e^{#1}}

	\newcommand{\comm}[2]{\left[#1,#2 \right]}
	
	\newcommand{\dd}{\mathbf{d}}

	\newcommand{\xxi}{\boldsymbol{\xi}}
	\newcommand{\abs}[1]{\left| #1 \right|}

	\newcommand{\braoket}[3]{\left< #1 \left| #2 \right| #3 \right>}

\begin{document}
\raggedbottom

\title{Resonantly hybridised excitons in moir\'e superlattices in van der Waals heterostructures}

\author{Evgeny M. Alexeev}
\affiliation{Department of Physics and Astronomy, University of Sheffield, Sheffield S3 7RH, UK}
\author{David A. Ruiz--Tijerina}
\affiliation{School of Physics and Astronomy, University of Manchester, Oxford Road, Manchester M13 9PL, UK}
\affiliation{National Graphene Institute, University of Manchester, Booth Street East, Manchester M13 9PL, UK}
\affiliation{Centro de Nanociencias y Nanotecnolog\'ia, Universidad Nacional Aut\'onoma de M\'exico, Apdo.\ Postal 14, 22800 Ensenada, Baja California, M\'exico}
\author{Mark Danovich}
\author{Matthew J. Hamer}
\author{Daniel J. Terry}
\affiliation{School of Physics and Astronomy, University of Manchester, Oxford Road, Manchester M13 9PL, UK}
\affiliation{National Graphene Institute, University of Manchester, Booth Street East, Manchester M13 9PL, UK}
\author{Pramoda K. Nayak}
\affiliation{Department of Physics, Indian Institute of Technology Madras, Chennai 600036, India}
\affiliation{Department of Energy Engineering and Department of Chemistry, Ulsan National Institute of Science and Technology (UNIST), 50 UNIST-gil, Ulsan 44919, Republic of Korea}
\author{Seongjoon Ahn}
\affiliation{Department of Energy Engineering and Department of Chemistry, Ulsan National Institute of Science and Technology (UNIST), 50 UNIST-gil, Ulsan 44919, Republic of Korea}
\author{Sangyeon Pak}
\affiliation{Department of Engineering Science, University of Oxford, Oxford OX1 3PJ, UK}
\author{Juwon Lee}
\affiliation{Department of Engineering Science, University of Oxford, Oxford OX1 3PJ, UK}
\author{Jung Inn Sohn}
\affiliation{Department of Engineering Science, University of Oxford, Oxford OX1 3PJ, UK}
\affiliation{Division of Physics and Semiconductor Science, Dongguk University-Seoul, Seoul 04620, South Korea}
\author{Maciej R. Molas}
\affiliation{Institute of Experimental Physics, Faculty of Physics, University of Warsaw, ul. Pasteura 5, 02-093 Warszawa, Poland}
\affiliation{National Graphene Institute, University of Manchester, Booth Street East, Manchester M13 9PL, UK}
\author{Maciej Koperski}
\affiliation{School of Physics and Astronomy, University of Manchester, Oxford Road, Manchester M13 9PL, UK}
\affiliation{National Graphene Institute, University of Manchester, Booth Street East, Manchester M13 9PL, UK}
\author{Kenji Watanabe}
\author{Takashi Taniguchi}
\affiliation{National Institute for Materials Science, 1-1 Namiki, Tsukuba 305-0044, Japan}
\author{Kostya S. Novoselov}
\affiliation{School of Physics and Astronomy, University of Manchester, Oxford Road, Manchester M13 9PL, UK}
\affiliation{National Graphene Institute, University of Manchester, Booth Street East, Manchester M13 9PL, UK}
\author{Roman V. Gorbachev}
\affiliation{Henry Royce Institute for Advanced Materials, University of Manchester, Manchester M13 9PL, UK}
\affiliation{School of Physics and Astronomy, University of Manchester, Oxford Road, Manchester M13 9PL, UK}
\affiliation{National Graphene Institute, University of Manchester, Booth Street East, Manchester M13 9PL, UK}
\author{Hyeon Suk Shin}
\affiliation{Department of Energy Engineering and Department of Chemistry, Ulsan National Institute of Science and Technology (UNIST), 50 UNIST-gil, Ulsan 44919, Republic of Korea}
\author{Vladimir I. Fal'ko}
\thanks{Corresponding author: vladimir.falko@manchester.ac.uk}
\affiliation{School of Physics and Astronomy, University of Manchester, Oxford Road, Manchester M13 9PL, UK}
\email{Corresponding author: vladimir.falko@manchester.ac.uk}
\affiliation{National Graphene Institute, University of Manchester, Booth Street East, Manchester M13 9PL, UK}
\affiliation{Henry Royce Institute for Advanced Materials, University of Manchester, Manchester M13 9PL, UK}
\author{Alexander I. Tartakovskii}
\thanks{Corresponding author: a.tartakovskii@sheffield.ac.uk}
\affiliation{Department of Physics and Astronomy, University of Sheffield, Sheffield S3 7RH, UK}
\email{Corresponding author: a.tartakovskii@sheffield.ac.uk}

\maketitle

{\bf Atomically-thin layers of two-dimensional materials can be assembled in vertical stacks held together by relatively weak van der Waals forces, allowing for coupling between monolayer crystals with incommensurate lattices and arbitrary mutual rotation \cite{GeimNature2013,Novoselov2016}. A profound consequence of using these degrees of freedom is the emergence of an overarching periodicity in the local atomic registry of the constituent crystal structures, known as a moir\'e superlattice \cite{kuwabara_1990}. Its presence in graphene/hexagonal boron nitride (hBN) structures \cite{Yankowitz2012} 
led to the observation of electronic minibands \cite{Dean2013,Hunt2013,Mishchenko2014a}, whereas its effect enhanced by interlayer resonant conditions in twisted graphene bilayers culminated in the observation of the superconductor-insulator transition at magic twist angles \cite{Cao2018}. Here, we demonstrate that, in semiconducting heterostructures built of incommensurate MoSe$_2$ and WS$_2$ monolayers, excitonic bands can hybridise, resulting in the resonant enhancement of the moir\'e superlattice effects. MoSe$_2$ and WS$_2$ are specifically chosen for the near degeneracy of their conduction band edges to promote the hybridisation of intra- and interlayer excitons, which manifests itself through a pronounced exciton energy shift as a periodic function of the interlayer rotation angle. This occurs as hybridised excitons (hX) are formed by holes residing in MoSe$_2$ bound to a twist-dependent superposition of electron states in the adjacent monolayers. For heterostructures with almost aligned pairs of monolayer crystals, resonant mixing of the electron states leads to pronounced effects of the heterostructure's geometrical moir\'e pattern on the hX dispersion and optical spectrum. Our findings underpin novel strategies for band-structure engineering in semiconductor devices based on van der Waals heterostructures\cite{Withers2015}.} 

Van der Waals heterostructures built from monolayers of transition-metal dichalcogenides (TMDs) present a unique combination of strong interaction with light, fast interlayer charge transfer and valley-dependent optical selection rules \cite{Rivera2018}, with potential for novel optoelectronic and valleytronic applications \cite{Rivera2018,Mak2016}. Despite significant advances in the understanding of moir\'e superlattices in graphene coupled to hBN \cite{Dean2013,Hunt2013,Mishchenko2014a} as well as in twisted graphene bilayers\cite{Cao2018}, for TMD heterobilayers the effects of this superlattice have been mostly explored theoretically\cite{hongyi_moire,mcdonald_intra}. Experimentally, it was found that the twist angle in a semiconducting heterobilayer influences the photoluminescence efficiency of interlayer excitons (bound electrons and holes situated in adjacent layers) \cite{Rivera2018,Heo2015,Nayak2017};
a variation in the photon emission energies of momentum- and spatially-indirect excitons has been reported in twisted MoS$_2$/WSe$_2$ heterobilayers\cite{Kunstmann2018a}, interpreted in terms of band structure variations caused by twist-angle dependent interlayer orbital hybridizations, similarly to previous reports on twisted MoS$_2$ bilayers \cite{Liu2014,VanDerZande2014}; and the most recent experiments have reported exciton localization in moir\'e potential wells in heterobilayers of TMDs with large band edge detuning \cite{MacDonald_moireX, Xu_moireX}.

Here, we report on the interplay between moir\'e superlattice effects and twist-controlled hybridisation of intralayer and interlayer excitons in MoSe$_2$/WS$_2$ heterobilayers. Close to perfect crystal alignment, we observe resonant enhancement of the hybridisation strength due to the nearly degenerate conduction band edges of these two monolayer semiconductors \cite{Kozawa2015} (  {Fig.~1a}). By studying a large number of heterobilayers with various interlayer orientations, fabricated from monolayers grown by chemical vapor deposition (CVD), as well as high-quality mechanically exfoliated heterostructures, we show that the hybridisation strength can be tuned continuously with the twist angle. This is manifested by a twist-dependent redshift of up to 60~meV of the photoluminescence (PL) peaks of the hybridized exciton (hX) states. Furthermore, additional features occur in the reflectance contrast (RC) spectra for near perfect alignment between the crystals, which we interpret as moir\'e superlattice effects leading to formation of minibands for hX excitons.

We have performed PL and RC studies over a wide range of temperatures on more than a 100 CVD and 5 mechanically exfoliated heterobilayers with various twist angles $\theta$ (defined in Fig.~1b).   {Figure 1} shows PL images \cite{Alexeev2017} of MoSe$_2$/WS$_2$ heterostructures assembled from mechanically exfoliated (c) and CVD-grown (d) monolayers.   {Fig.~2a} compares room-temperature PL spectra of CVD-grown MoSe$_2$ and WS$_2$ monolayers, and a MoSe$_2$/WS$_2$ structure with $\theta=2^\circ$. In the heterobilayer region, at energies close to the MoSe$_2$ and WS$_2$ intralayer A-exciton PL peaks (see Fig.~1a for the origin of the main optical transitions), P1 and P2 PL peaks are observed. P2 appears at the same energy as the monolayer WS$_2$ PL line ($\approx 1.96\,{\rm eV}$), due to the strong signal collected from the surrounding monolayer WS$_2$ (see   {Fig.~1d}). By contrast, P1 displays a $57\,{\rm meV}$ red shift with respect to the MoSe$_2$ A exciton line at 1.573 eV. 

  {Fig.~2b} shows normalized PL spectra for the P1 peak for heterobilayers with twist angles approaching $0^\circ$ (red) and $60^\circ$ (blue). Strikingly, the P1 peak energy depends strongly on the twist angle, decreasing continuously as the stacking approaches lattice alignment ($\theta=0^\circ$), or anti-alignment ($\theta=60^\circ$). This is summarized in   {Fig.~2c}, showing two distinct trends: a steep variation of $\approx 60\,{\rm meV}$ for $\theta$ close to $0^\circ$ and $60^\circ$, and a plateau at around 1.56 eV for large misalignment angles $8^\circ < \theta < 52^\circ$. The P1 integrated intensity and linewidth vary gradually as $\theta$ departs from $30^\circ$ (  {Extended Data Figure~1a}). The linewidth displays an additional sharp increase for twist angles close to $0^\circ$ and $60^\circ$, similarly to the observed variation of the peak energy (  {Extended Data Figure~1b}).

By contrast to the $120^\circ$ periodicity of the lattice and flake symmetry, the sharp variation of the P1 line energy and width with a $\theta\approx 60^\circ$ period strongly suggests a connection to the misalignment between the monolayer Brillouin zones (BZs) produced by the twist, which introduces a momentum mismatch between the conduction and valence band edges of the two monolayers, located at their $K$ and $K'$ valleys (  {Fig.~3a-b}) \cite{Zhang2017}. As shown in   {Fig.~3a}, for perfect lattice alignment (anti-alignment), the valley mismatch $\Delta\KK=\KK_{\text{WS}_2}-\KK_{\text{MoSe}_2}$ ($\Delta\KK'=\KK_{\text{WS}_2}'-\KK_{\text{MoSe}_2}$) is minimized. In these configurations the MoSe$_2$ A and B excitons are brought into close momentum-space proximity with interlayer excitons formed by MoSe${}_2$ holes and WS${}_2$ electrons, allowing them to hybridise through interlayer conduction-band tunnelling. Although the interlayer tunnelling strength has been estimated to be small for states at the BZ corners \cite{Wang2016b} for TMD bilayers, the near resonance of the MoSe${}_2$ and WS${}_2$ conduction-band edges \cite{abinitio_1, abinitio_2}%, abinitio_3} 
promotes hybridisation and the formation of hXs. This is similar to interlayer hybridization effects recently predicted \cite{Deilmann2018,Gerber2018} and experimentally observed \cite{Gerber2018,Slobodeniuk2018} in TMD homobilayers.   {We thus interpret the P1 line shift in   {Fig.~2} in terms of the twist-angle-dependent hX band structure. Furthermore, our theory reveals a strong influence of the moir\'e superlattice on the twist-angle dependent energies of bright hXs for twist angles near $0^\circ$ and $60^\circ$ (see   {Supplementary Notes 1 and 2} and Extended Data   {Figs.~2-4}), which is beyond the scope of previous descriptions based on harmonic moir\'e potentials \cite{hongyi_moire,mcdonald_intra}.}

The two lowest-energy bright interlayer excitons involving same-spin conduction and valence bands of opposite layers are shown in   {Fig.~3a}, labeled iX and iX', respectively. In general, at the bottom of their dispersions, iX and iX' are expected to be ``momentum-dark,'' due to the rotational misalignment   {(Fig.~3a)}. However, for closely-aligned or anti-aligned structures, momentum-space proximity with the intralayer MoSe$_2$ A exciton allows them to hybridise, given their shared hole band. The resulting hXs pick up oscillator strength from their X$_{\rm A}$ component \cite{hX_theory}, and their energies are sharply modulated for twist angles close to $0^\circ$ and $60^\circ$. Analogously, MoSe${}_2$ B excitons (X$_{\rm B}$) can hybridise with iXs formed by carriers in the corresponding spin states. Note that spin-splitting and spin-valley locking of iXs lead to different energies and compositions of the hX states for $\theta = 0^\circ$ and $60^\circ$, producing the asymmetry seen in   {Fig.~2c}.

The MoSe$_2$ and WS$_2$ conduction band states have plane-wave projections onto all Bragg vectors of their corresponding reciprocal lattices, $\GG_n^{\rm MoSe_2}$ and $\GG_n^{\rm WS_2}$ (red and green arrows in   {Fig.~3b}). As a result, momentum conservation allows interlayer tunnelling between first BZ wave vectors of the two layers satisfying $\kk_{\rm WS_2}-\kk_{\rm MoSe_2} = \Delta\KK^{(')} + \bb_n$, where $\kk_{\rm MoSe_2}$ and $\kk_{\rm WS_2}$ are measured from the respective $K$ or $K'$ valley, and $\bb_n=\GG_n^{\rm WS_2}-\GG_n^{\rm MoSe_2}$ are the first Bragg vectors of the moir\'e superlattice, which define the twist-angle-dependent mini Brillouin zone (mBZ) shown in purple in   {Fig.~3b} (see also   {Extended Data Fig.~3}). 

The emergent periodicity leads to folding of the intra- and interlayer exciton bands into the mBZ, forming moir\'e mini bands, hybridised into hXs by interlayer electron tunnelling. Theoretically, this becomes evident in the twist-angle-dependent behaviour of the hX bands, which we calculate for $0^\circ \le \theta \le 10^\circ$ along the momentum path $\overline{-\Delta\KK,\mathbf{\Gamma},\Delta\KK}$ (dotted line in   {Fig.~3b}), with $\mathbf{\Gamma}$ representing zero center-of-mass exciton momentum: as shown in   {Fig.~3c}, a continuous energy drop of the lowest optically-active $\mathbf{\Gamma}$-hX as $\theta \rightarrow 0^\circ$ is observed. Experimentally, this behaviour is observed in PL as the P1 line red-shift of   {Figs.~2b,c}, and a similar situation arises near $60^\circ$, where the hX is formed by iX' (  {Extended Data Fig.~4}). The calculated twist-angle dependence of the $\mathbf{\Gamma}$-hX energy is shown by the blue curve in   {Fig.~2c}, obtained using a difference of $13\,{\rm meV}$ between the X$_{\rm A}$ and iX energies (  {Fig.~3c}), an electron hopping strength of $26\,{\rm meV}$, and setting the X$_{\rm A}$ energy at room temperature to $1.555\,{\rm eV}$, as obtained from the $8^\circ < \theta < 52^\circ$ plateau in the experimental data   {(Extended Data Table~I)}. Furthermore, we attribute the sharp linewidth variation with twist angle for $\theta$ close to $0^\circ$ and $60^\circ$, shown in   {Extended Data Fig.~1b}, to the out-of-plane electric dipole moment of the hX state, inherited from its iX component, which makes the hX energy susceptible to random electric fields (see   {Supplementary Note~3}).

  {Figure~3e} shows calculated absorption spectra for the heterobilayer as a function of twist angle, predicting three-peak structures close to the X$_{\rm A}$ and X$_{\rm B}$ energies for twist angles close to $0^\circ$.   {In this case, we have used an X${}_{\rm A}$ energy of $1.632\,{\rm eV}$, corresponding to our exfoliated samples (see below), and kept the same X-iX detuning and hopping strength obtained from the CVD data} (  {see Extended Data Table I}). The peaks labeled hX${}_1$ and hX${}_2$ are two hX states resulting from resonant hybridisation of the X$_{\rm A}$ band with the lowest-energy iX band at the $\Gamma$ point. The energy shift of hX${}_1$ with twist angle indicates that it corresponds to P1 in our PL measurements, whereas the transfer of the oscillator strength from hX${}_1$ to hX${}_2$ for $\theta \rightarrow 0^\circ$ shows enhanced mixing of the former with the $\boldsymbol{\Gamma}$-iX state. By contrast, the hX${}_3$ peak originates from the first moir\'e mini band, formed by folding of the X${}_{\rm A}$ band onto the mBZ. This is illustrated in   {Fig.~3e}, where we have set the X${}_{\rm A}$-iX tunnelling strength to zero, such that the red (blue) bands represent pure X${}_{\rm A}$ (iX) states. The black arrow indicates the point where the first X$_{\rm A}$ moir\'e mini bands cross the $\Gamma$ point. For non-zero tunnelling, these mini bands will mix with the bright $\boldsymbol{\Gamma}$-X$_{\rm A}$ state below. In other words, hX${}_3$ becomes semi-bright through moir\'e umklapp processes that mediate the exciton-photon interaction \cite{hX_theory}, and constitutes a direct signature of the moir\'e effects on the exciton band structure. An analogous three-line structure appears in the calculated spectrum near X$_{\rm B}$, around $1.8$ eV in   {Fig.~3d}.

We carried out RC measurements of the same CVD-grown heterostructures at low temperature, where the spectral linewidths are narrower.   {Fig.~2d} shows the RC spectra measured at $T$=10K for several CVD MoSe$_2$/WS$_2$ heterobilayers and a MoSe$_2$ monolayer. Similarly to room-temperature PL, we observe in   {Fig.~2d} a 20 meV red-shift of the X${}_{\rm A}$ exciton resonance in the misaligned heterobilayer (orange) for $\theta$=31$^\circ$, as compared to the uncoupled monolayer (black). The strong red-shift of the RC peaks and significant reduction of their intensities observed for $\theta$ close to $0^\circ$ and $60^\circ$ is consistent with our room-temperature PL measurements. In addition, note the appearance of weak new features in the spectra for $\theta = 1^\circ$ and $59^\circ$, within 50 meV above the main reflectance feature (  {Fig.~2d} and   {Extended Data Fig.~5a}). We associate these spectral signatures with the absorption peaks hX$_{1,2,3}$ in   {Fig.~3d}. Similar twist-angle-dependent red-shifts are also observed in the vicinity of the X$_{\rm B}$ energy, as shown in   {Extended Data Fig.~5b}, where also a clear reduction of the intensity of the RC peaks is evident for $\theta$ approaching $0^\circ$ and $60^\circ$, consistent with   {Fig.~3d}. However, no additional features are distinguishable, due to the broader X${}_{\rm B}$ linewidth.

A dramatic improvement in both RC and PL spectral resolution is observed in the high-quality MoSe$_2$/WS$_2$ heterostructures, assembled in a glove-box from mechanically exfoliated monolayers and fully encapsulated in hBN.   {Figure~4a} shows a comparison of normalised PL spectra measured for a monolayer MoSe$_2$ (black), and MoSe$_2$/WS$_2$ heterobilayers with $\theta$ of 12$^\circ$ (orange) and $\approx 1.8^\circ$ (red). The black and red spectra are measured in a single sample, where in addition to the heterobilayer, a monolayer MoSe${}_2$ region is also present. The PL spectrum for the monolayer area shows a very pronounced trion (charged exciton) peak, labeled X* (with intensity exceeding 120000 counts/s/$\mu W$), and a weaker neutral exciton, X${}_{\rm A}$. In the structure with misaligned MoSe$_2$ and WS$_2$ monolayers, both peaks red-shift by around 20 meV (orange spectrum). In this sample, MoSe$_2$ completely overlaps with WS$_2$, and thus no PL from uncoupled MoSe$_2$ is observed. For the high-quality aligned heterobilayer ($\theta\approx 1.8^\circ$, red spectrum), the PL spectrum shows 4 peaks with a maximum intensity of 1500 counts/s/$\mu W$. As can be deduced from the identical spectral positions, the two peaks at higher energies correspond to PL collected from the remote MoSe$_2$ monolayer area. The other two features, labeled hX* and hX$_1$, originate in the heterostructure region and exhibit a strong red shift of 38 and 32 meV from X* and X${}_{\rm A}$ in the $\theta\approx 12^\circ$ structure, respectively, consistent with the behaviour observed in the CVD samples. We observed similar PL spectra in other high-quality exfoliated samples   {(Extended Data Fig.~6e)}.
 
The green curve in   {Fig.~4b} shows the RC spectrum of the exfoliated heterobilayer with $\theta=12^\circ$,   {where a single strong resonance corresponding to the X${}_{\rm A}$ transition is observed}. In stark contrast, four features are clearly observed in the RC spectrum of the aligned sample (  {Fig.~4c} and   {Extended Data Fig.~6b-d}), labeled as hX*, hX$_1$, hX$_2$ and hX$_3$. hX* and hX$_1$ directly correspond to the identically labeled peaks in PL, whereas the new features hX$_2$ and hX$_3$ are unrelated to X* and X${}_{\rm A}$ peaks observed in the monolayer MoSe$_2$ PL spectrum. We attribute hX${}_2$ and hX${}_3$ to the moir\'e miniband states of hybridised excitons labeled in   {Fig.~3d}. Temperature dependent studies in   { Fig.~5} show that all hX peaks persist in RC spectra up to 105 K, above which the observation of hX* and hX$_3$ becomes difficult due to broadening. We tentatively ascribe hX* to a recombination of hybridised excitons bound to an additional charge. At room temperature   {(Extended Fig.~7)}, the most pronounced peak in PL and RC is hX$_1$, supporting its identification with PL peak P1 in   {Fig.~2}. Cooling down to $T\approx 220$K is necessary for unambiguous identification of hX$_2$, as shown in   {Fig.~5}. On the other hand, while the PL trion peak X* vanishes at $T\ge 65$K, a new PL feature corresponding to hX$_2$ appears for $T>100$K due to thermal activation, in agreement with our calculations (see   {Supplementary Note~1} and   {Extended Data Fig.~8}). 

Moir\'e excitons have been previously discussed \cite{hongyi_moire,mcdonald_intra,MacDonald_moireX,Xu_moireX} in type-II TMD heterobilayers with strong detuning between the band edges of their constituting layers \cite{Zhang2018}. In those cases, interlayer hybridisation of carriers is weak, and the moir\'e effects are dominated by a periodic exciton potential produced by the band gap modulation along the heterostructure plane. Our results indicate that resonant interlayer hybridisation in MoSe${}_2$/WS${}_2$, such as studied here, is the dominant mechanism driving moir\'e superlattice effects, producing hX states that inherit spectral properties of the intralayer and interlayer excitons. Our PL and RC measurements for varying twist angle in both CVD and exfoliated samples provide strong evidence for the appearance of semi-bright hybridised exciton states, and in particular for spectral features originating from higher hybridised moir\'e exciton mini bands, enabled by umklapp exciton-photon processes mediated by the moir\'e superlattice. We anticipate a similar effect of the hybridised exciton formation and resonant enhancement of moir\'e superlattice effects in a broader class of 2D material heterostructures with close conduction or valence band alignment. In such materials, the twist-angle-controlled resonant effects reported in this work constitute an unprecedented approach for carrier and exciton band structure engineering.

\section{References}
\bibliographystyle{naturemag}
%\bibliography{library}

\section{Acknowledgments}

The authors thank the financial support of the European Graphene Flagship Project under grant agreement 696656, EC Project 2D-SIPC, and EPSRC grant EP/P026850/1. E.M.A.\ and A.I.T.\ acknowledge support from EPSRC grants EP/M012727/1 and the European Union's Horizon 2020 research and innovation programme under 
ITN Spin-NANO Marie Sklodowska-Curie grant agreement no. 676108. D.A.R-T.\ and V.I.F.\ acknowledge support from ERC Synergy Grant Hetero2D, EPSRC EP/N010345, and the Lloyd Register Foundation Nanotechnology grant. K.S.N.\ thanks financial support from the Royal Society, EPSRC, US Army Research Office and ERC Grant Hetero2D. H.S.S.\ acknowledge the research fund (NRF-2017R1E1A1A01074493) by National Research Foundation by the ministry of Science and ICT, Korea. M.R.M.\ acknowledges support from the National Science Centre (no. UMO-2017/24/C/ST3/00119). R.V.G.\ acknowledges financial support from the Royal Society Fellowship Scheme and EPSRC CDT Graphene-NOWNANO EP/L01548X. K.W. and T.T. acknowledge support from the Elemental Strategy Initiative
conducted by the MEXT, Japan and the CREST (JPMJCR15F3), JST.

\section{Author contributions}

E.M.A.\ carried out microscopy and optical spectroscopy experiments. E.M.A.\ and A.I.T.\ analyzed optical spectroscopy data. D.A.R.-T., M.D., V.I.F.\ developed the theory. P.K.N., S.A., S.P, J.L., J.I.S. and H.S.S. carried out CVD growth of the monolayers and fabricated the heterobilayer samples. M.J.H., D.J.T. and R.G. fabricated mechanically exfoliated samples using glove box techniques. M.R.M. and M.K. carried out SHG measurements on exfoliated samples. K. W. and T. T. synthesized the hBN crystals. E.M.A., D. A.R.-T., A.I.T.\ and V.I.F. wrote the manuscript. E.M.A., K.N.S., V.I.F.\ and A. I. T.\ conceived the experiment. All authors participated in discussions. A. I. T. oversaw the project.

\section{Author Information}
Correspondence and requests for materials should be addressed to Alexander Tartakovskii at a.tartakovskii@sheffield.ac.uk and Vladimir Fal'ko at Vladimir.Falko@manchester.ac.uk.

\section{Figure captions}

\label{fig:SampleImage}
\textbf{Figure 1. Twisted MoSe$_2$/WS$_2$ heterobilayers.} 
(a) Schematic band diagram of MoSe$_2$/WS$_2$ showing staggered alignment of the valence bands (VB) and the near degeneracy of the conduction band (CB) edges. The carrier spin orientations are labeled, and the spin-orbit splittings are schematically shown as energetically separated valence and conduction sub-bands. CB and VB states involved in the formation of the bright A and B excitons to are connected by arrows. The dashed arrows represent resonant hybridisation between conduction band states of the two layers. (b) Real--space configuration of the heterobilayer. Mo (W) atoms are shown in red (green), and all chalcogens are depicted in orange. The angle $\theta$ between same lattice vectors of the two layers, e.g.\ $\mathbf{a}_{1}^{{\rm MoSe}_2}$ and $\mathbf{a}_{1}^{{\rm WS}_2}$, translates into an angle $\theta$ between the reciprocal lattice vectors, resulting in misalignment of the two Brillouin zones.
(c) PL image of one of the studied mechanically exfoliated hBN/MoSe$_2$/WS$_2$/hBN samples acquired using a modified optical microscope (see Methods and Ref.\ \cite{Alexeev2017} for details). Orange and pink correspond to PL emitted from WS$_2$ and MoSe$_2$ regions, respectively. Edges of the monolayer flakes are marked. Dark (low PL) areas correspond to the heterobilayer region where the WS$_2$ and MoSe$_2$ intralayer A exciton PL is quenched. (d) PL image of one of the studied heterobilayers assembled from CVD-grown monolayers, where red and pink correspond to PL from WS$_2$ and MoSe$_2$, respectively. The scale bar in (c) and (d) corresponds to 10 $\mu$m. \\

\label{fig:TAngleDependence}
	\textbf{Figure 2. Optical properties of twisted MoSe$_2$/WS$_2$ heterobilayers.} (a) Room-temperature PL spectra measured in CVD-grown MoSe$_2$ (black) and WS$_2$ (magenta) monolayers, as well as in a mechanically stacked MoSe$_2$/WS$_2$ heterostructure with rotation angle $\theta=2^{\circ}$ between the layers (blue). The two PL peaks appearing in the heterobilayer are labeled P1 and P2. (b) Normalized PL spectra showing peak P1 from (a) acquired in MoSe$_2$/WS$_2$ heterobilayers with interlayer twist angles $\theta$ ranging from 1 to 59$^\circ$, indicated above each curve. A typical PL spectrum for a monolayer MoSe$_2$ is shown with a dashed black curve. The MoSe$_2$ A exciton PL peak is labeled X${}_{\rm A}$, and its energy is marked with the vertical dashed line. (c) Variation of the P1 PL peak energy with twist angle measured on two substrates containing MoSe$_2$/WS$_2$ heterobilayers. Data for individual substrates are shown with light and dark red symbols. See description of the fitting procedure in   {Methods}. The blue curve shows results of theoretical calculations as described in the text and   {Supplementary Note~1}. The dashed horizontal line shows the spectral position of X${}_{\rm A}$ in an isolated MoSe$_2$ monolayer. (d) Low temperature (T=10 K) reflectance contrast (RC) spectra measured near the A exciton energy in an isolated MoSe$_2$ monolayer (black), and in MoSe$_2$/WS$_2$ heterostructures with various twist angles $\theta$ (with values shown next to each curves). See explanation of the RC experiment, in particular parameters $R$ and $R_0$, in Methods. The A exciton feature in the isolated MoSe$_2$ monolayer spectrum is labeled X${}_{\rm A}$. Arrows indicate new RC features appearing in the spectra for structures with nearly perfect (anti-)alignment. Vertical lines show the positions of the maximum derivative of the X${}_{\rm A}$ feature in the monolayer (black) and misaligned heterobilayer (orange) spectra.\\

\label{fig:line_fitting}
{\bf Figure 3. Theoretical moir\'e bands and absorption spectrum of hybridised excitons.} (a) MoSe$_2$ and WS$_2$ electronic band structures and BZ alignment for twist angle $\theta$ (the conduction band spin-orbit couplings $\Delta_{\rm SO}$ and $\Delta_{\rm SO}'$ are exaggerated for clarity). Spin-down (spin-up) bands are coloured red or green (grey). The wavy red line represents the formation of the zero-momentum MoSe$_2$ A exciton, X${}_{\rm A}$. The two interlayer exciton states iX and iX', involving WS$_2$ electrons of opposite valleys, are shown with wavy blue lines. (b) Moir\'e mini Brillouin zone (mBZ, purple) defined by the moir\'e Bragg vectors $\bb_n = \GG_n^{\rm WS_2}-\GG_n^{\rm MoSe_2}$ (purple arrows). 
(c) hybridised exciton (hX) bands along the mBZ path defined in (b) for twist angles $\theta=0^\circ,\,2^\circ$ and $10^\circ$ (for angles close to $60^\circ$, see Extended Data Fig.~4). For reference, the decoupled intra- and interlayer (iX) exciton bands are shown in red and blue dashed curves, respectively. Optically active hybridised exciton states in that energy range are marked with yellow dots, whereas unmarked $\Gamma$ excitons are of iX nature. (d) Absorption spectrum as a function of twist angle. The MoSe$_2$ A and B exciton resonances X$_{\rm A}$ and X${}_{\rm B}$ are indicated for large twist angles, where hybridisation effects become negligible. By contrast, the three lines labeled hX${}_{1,2,3}$ appearing for $\theta\approx 0^\circ$ correspond to bright resonantly hybridised excitons in the range of MoSe${}_2$ X${}_{\rm A}$; analogous features in the range of X${}_{\rm B}$ are not labeled. In particular, hX${}_3$ originates from hybridisation of the first folding of the X${}_{\rm A}$ band into the mBZ with the bright zero-momentum X${}_{\rm A}$ state, indirectly through the iX band. This is illustrated in (e), where we show the X${}_{\rm A}$ (red) and iX (blue) bands in the absence of hybridisation, for clarity. The black arrow indicates the point where the first folded X${}_{\rm A}$ bands cross the $\Gamma$ point. Thus, hX${}_3$ constitutes a direct signature of the moir\'e superlattice effect.\\	

	\label{fig:NewLowTDataExfoliated}
	\textbf{Figure 4. Low temperature (T=10 K) optical spectra for exfoliated MoSe$_2$/WS$_2$ heterobilayers and MoSe$_2$ monolayers.} (a) PL spectra of the MoSe$_2$ monolayer (black), misaligned MoSe$_2$/WS$_2$ heterobilayer with $\theta=12^\circ$ (orange), and the aligned MoSe$_2$/WS$_2$ heterobilayer with $\theta\approx 1.8^\circ$ (red). All samples are encapsulated in hBN. Black and red spectra are measured in the same sample, where an uncoupled MoSe$_2$ monolayer was present. Vertical lines mark the positions of A exciton (X${}_{\rm A}$) and trion (X*) resonances in the uncoupled MoSe$_2$ monolayer, also visible in the red spectrum due to the monolayer proximity and its much stronger PL (see text and   {Extended Data Fig.~6}). Peak hX$_1$ appears in the PL spectrum of the $\theta\approx1.8^\circ$ sample, as predicted in Fig.~3d, together with an additional interlayer exciton line hX${}^*$ (see text). (b) PL (orange) and RC (green) spectra for the misaligned MoSe$_2$/WS$_2$ heterobilayer, with $\theta=12^\circ$. X${}_{\rm A}$ labels the A exciton RC and PL features. (c) PL (red) and RC (green) spectra for the aligned MoSe$_2$/WS$_2$ heterobilayer, with $\theta\approx 1.8^\circ$. In addition to hX${}_1$ and hX${}^*$, two other features, hX${}_2$ and hX${}_3$, appear in the RC spectrum, in agreement with   {Fig.~3d}.\\
	
\label{fig:NewTdepData}
	\textbf{Figure 5. Temperature-dependent PL and RC spectra for aligned MoSe$_2$/WS$_2$ heterobilayer.} The data are shown for the same sample with hBN-encapsulated MoSe$_2$/WS$_2$ heterobilayer with $\theta\approx 1.8^\circ$, for which the low temperature data are presented in Fig.~4c. Here, PL and RC spectra are shown in black and green, respectively. Similarly to Fig.~4c, RC and PL peaks are labeled hX* and hX$_{1-3}$. X* and X${}_{\rm A}$ mark PL peaks from the remote isolated MoSe$_2$ monolayer. The bottom panel presents the same low temperature data as shown in Fig.4c.

\section{Methods}
\textbf{Fabrication of twisted MoSe$_2$-WS$_2$ heterobilayers}\label{sec:growth}

Triangle-shaped monolayers of MoSe${}_2$ and WS${}_2$ were produced by chemical vapor deposition (CVD) growth. Single--layer MoSe${}_2$ crystals were grown on a c-plane sapphire substrate by the vaporization of MoO$_3$ and Se powders in a 2-inch quartz tube furnace in a controlled gaseous environment. A typical run consisted of loading 60 mg of MoO$_3$ source powder and 100 mg of Se powder into two ceramic boats, which were then placed into the center heating zone and upwind the low temperature zone in the same quartz tube. A piece of sapphire substrate, acting as a deposition acceptor, was placed downstream adjacent to the MoO$_3$ powder. The temperature of MoO$_3$ powder was raised to 600$^{\circ}$ C at a rate of 25$^{\circ}$C/min and then increased to 700$^{\circ}$ C within 10 min. At the same time, the temperature of the Se powder was raised to 275$^{\circ}$ C using an external heating coil. After reaching their target values, the temperatures of MoO$_3$ and Se were maintained at the same levels to facilitate the MoSe${}_2$ growth. The vapor-phase reactants were transported by Ar gas flow (60 sccm), and selenization was carried out by the flowing H${}_2$ reductant gas (12 sccm), thereby facilitating the growth of the 2D MoSe${}_2$ crystals at the growth region. At the end of the growth, the furnace was fast cooled to room temperature in a pure Ar atmosphere.

For WS${}_2$ growth, 50 mg of WO$_3$ powder were placed at the center of an alumina boat, and a silicon substrate with a 300 nm-thick SiO${}_2$ layer was placed upside down above it. Another alumina boat containing 200 mg of sulfur powder was placed upstream of the furnace. The furnace was heated to 950$^{\circ}$C and the temperature was maintained for 20 minutes with a flow of Ar gas (150 sccm) in order to grow monolayer WS${}_2$. The furnace was then allowed to cool down naturally.

For twisted heterobilayer fabrication, the monolayer MoSe${}_2$ flakes grown on a sapphire substrate were transferred onto a 300 nm SiO${}_2$/Si substrate using a poly(methyl methacrylate) (PMMA) transfer process. Few drops of PMMA (950 K) were spin-coated on the MoSe${}_2$/sapphire substrate at 4000 RPM for 60 second and backed at 120$^{\circ}$ C for 30 min. To facilitate the separation of the PMMA membrane, a drop of deionized water was put at the interface between PMMA and sapphire substrate. The PMMA-coated MoSe${}_2$ monolayers were then peeled off the substrate using sharp tweezers and transferred into deionized water. The floating PMMA/MoSe${}_2$ membrane was transferred onto a prepared WS${}_2$/(SiO${}_2$/Si) substrate and dried at 80$^{\circ}$ C for few minute to evaporate the water at the interface. Finally, the PMMA layer was dissolved in acetone, and the substrate was cleaned using isopropyl alcohol. 

High-quality fully-encapsulated MoSe$_2$/WS$_2$ samples were fabricated using PMMA-assisted dry-peel transfer. In order to minimize contamination, heterostructures were fabricated using remotely controlled micro-manipulation setup placed inside an argon chamber with $<$0.1~ppm O$_2$ and H$_2$O. The bulk MoSe$_2$ and WS$_2$ crystals were mechanically exfoliated onto a 90~nm layer of PMMA coated on a silicon wafer. Monolayers were then identified via optical microscopy, as well as through luminescence imaging in the dark-field configuration for WS$_2$ \cite{Alexeev2017}. Those crystals which had adjacent straight edges at 0${}^\circ$, 60${}^\circ$ or 120${}^\circ$ to one another (indicating one of the crystallographic axes) were then selected and picked up onto an hBN film ($\lesssim$~10~nm thickness) held by a PMMA membrane. During the transfer of the second TMD layer the edges were aligned to within 2${}^\circ$ of the desired angle and finally transferred onto another hBN film ($\lesssim$~20~nm) exfoliated onto an oxidised silicon wafer (70~nm SiO$_2$) to achieve full encapsulation. In order to prevent spontaneous rotation of the TMDC layers and deterioration of MoSe$_2$ crystalline quality we avoided exposing the heterostructures to the temperatures above 70${}^\circ$~C.

\textbf{Twist angle measurements and spectral fitting}.

The interlayer twist angle in CVD MoSe${}_2$/WS${}_2$ heterobilayers was extracted from the microscope images by comparing the orientation of the two materials. 
It has been shown [35,36] that triangular transition-metal dichalcogenide (TMD) flakes have zigzag edges, which appear sharp or diffuse in microscopy images when the edges terminate with transition metal atoms or chalcogens, respectively. Therefore, as the sharp terminations of the flakes shown in Fig.~1d of the main text indicate transition-metal termination, the relative flake orientation directly corresponds to the twist angles between their lattice vectors, up to $120^\circ$ rotations given the $C_3$ rotational symmetry of both the flakes and their underlying lattices. Similarly, for flakes with the boundaries of different types, the interlayer twist angle is determined by subtracting 60${}^\circ$ from the relative rotation measured from the microscope image.

With these considerations, we were able to determine the relative twist angles $\theta$ between the different MoSe${}_2$ flakes and the bottom WS${}_2$ monolayer from the microscopy images with $\sim 1^\circ$ accuracy. Inherent to our method is an uncertainty in the angular measurement when the type of the flake boundary is not clear, and the edge may be formed by either transition metal or chalcogen atoms.

Although only a few of our samples displayed these characteristics, this uncertainty may be behind the symmetric shape of the data presented in   {Fig.~2c} and   {Extended Data Fig.~1b}.

To identify the relative twist angle of the exfoliated TMD monolayers, second harmonic generation (SHG) measurements were performed [37], see Extended Data Fig.9. All SHG measurements were taken at room temperature using a custom-built system with a Toptica FemtoFErb $\sim90$~fs SAM mode-locked laser with a repetition rate of 80~MHz centred at 785~nm. For each measurement, the laser light had a typical incident power of 500~$\mu$W, was linearly polarized and focused to a spot size of $<$ 2 $\mu$m by a 50x objective lens (NA = 0.36). The SHG signal was separated from the reflected light using of a beam splitter and short-pass filter. A second linear polarizer (analyzer) was placed in the SHG signal path and aligned parallel to the excitation polarization. A motorized half-wave plate was located above the objective and rotated in order to obtain angle resolution. The SHG signal for each measurement was then processed by a spectrometer with a grating of 300 groves per millimetre and exposed onto a liquid nitrogen cooled charge-coupled device for 5 seconds.

The position, width and integrated intensity of the hybridized exciton peak were extracted from the room-temperature PL spectra of the corresponding regions by fitting with a single Lorentzian peak. In several regions, however, the proximity to isolated monolayer areas or the presence of contamination trapped between the layers has led to the detection of an uncoupled MoSe${}_2$ signal, which appears as a high-energy shoulder in the PL spectrum (see top curve in   {Fig.~2b}). The PL spectra containing isolated MoSe${}_2$ signal were fitted with two Lorentzian peaks, and the parameters of the lower-energy peak were used for the twist angle dependence.

\textbf{Photoluminescence imaging, and photoluminescence and reflectance contrast spectroscopy.}

The photoluminescence (PL) images of the heterobilayer samples were acquired using a modified bright-field microscope (LV150N, Nikon) equipped with a color camera (DS-Vi1, Nikon). The near-infrared emission from the white light source was blocked with a 550~nm short-pass filter (FESH0550, Thorlabs), and a 600~nm long-pass filter (FELH0600, Thorlabs) was used to isolate the PL signal from the sample. The full description of the system is available in Ref.~\onlinecite{Alexeev2017}.

Spectrally--resolved PL and reflectance contrast (RC) measurements were performed using a custom-built micro-PL setup. For PL, the excitation light centred at 2.33 eV was generated by a diode-pumped solid-state laser (CW532-050, Roithner), while for RC a stabilized Tungsten-Halogen white-light source (SLS201L, Thorlabs) was used. The excitation light was focused onto the sample using a 50x objective lens (M Plan Apo 50X, Mitutoyo). The PL and RC signals collected in the backwards direction were detected by a 0.5~m spectrometer (SP-2-500i, Princeton Instruments) with a nitrogen cooled CCD camera (PyLoN:100BR, Princeton Instruments). The PL signal was isolated using a 550 nm short-pass filter (FELH0550, Thorlabs). The RC spectra were derived by comparing the spectra of white light reflected from the sample and the substrate as $RC(\lambda) = (R(\lambda) - R_0(\lambda))/(R(\lambda) + R_0(\lambda))$, where $R$ ($R_0$) is the intensity of light reflected by the sample (substrate). The room-temperature measurements were performed in ambient conditions. The low temperature measurements were carried out using a continuous flow liquid helium cryostat, where the sample was placed on a cold finger with the base temperature of 10 K.

\textbf{References for Methods}

[35] van der Zande, A. M. et al. Grains and grain boundaries in highly crystalline monolayer molybdenum disulphide. Nature Materials {\bf 12}, 554-561 (2013).

[36] Zhu, D. et al. Capture the growth kinetics of CVD growth of two-dimensional MoS2. npj 2D Materials and Applications {\bf 1}, 8 (2017).

[37] Hsu, W.-T. et al. Second harmonic generation from artificially stacked transition metal dichalcogenide twisted bilayers. ACS Nano {\bf 8}, 2951-2958 (2014).

\newpage

\section{Figures}

\begin{figure}[h!]
	\centering
	\includegraphics[width=1\columnwidth]{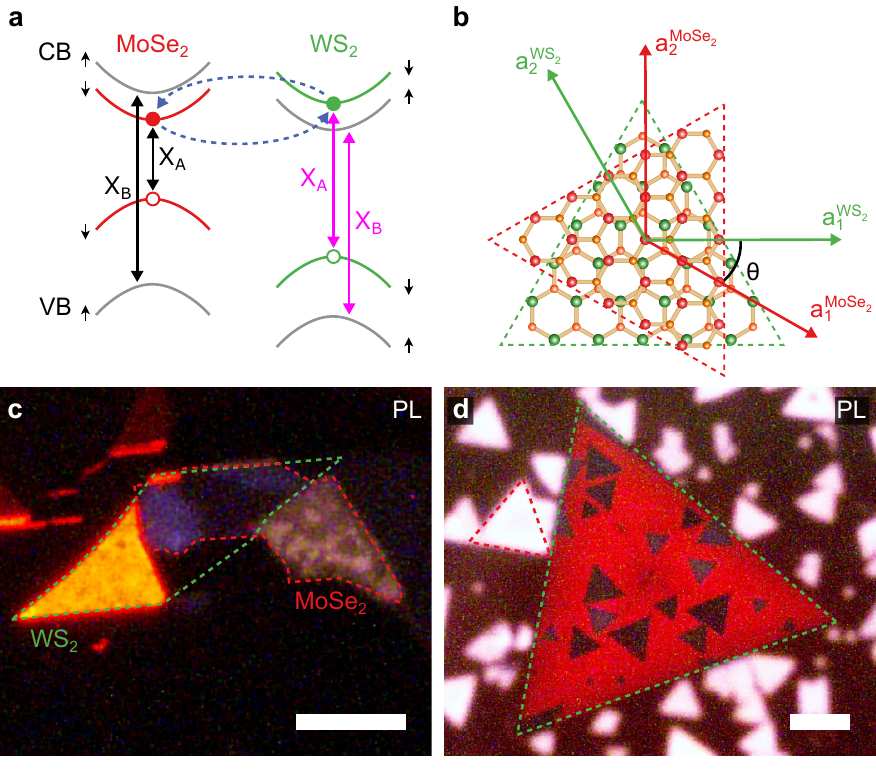}
	\caption{}
	\end{figure}
	
	\newpage

\begin{figure}[h!]
	\centering
	\includegraphics[width=1\columnwidth]{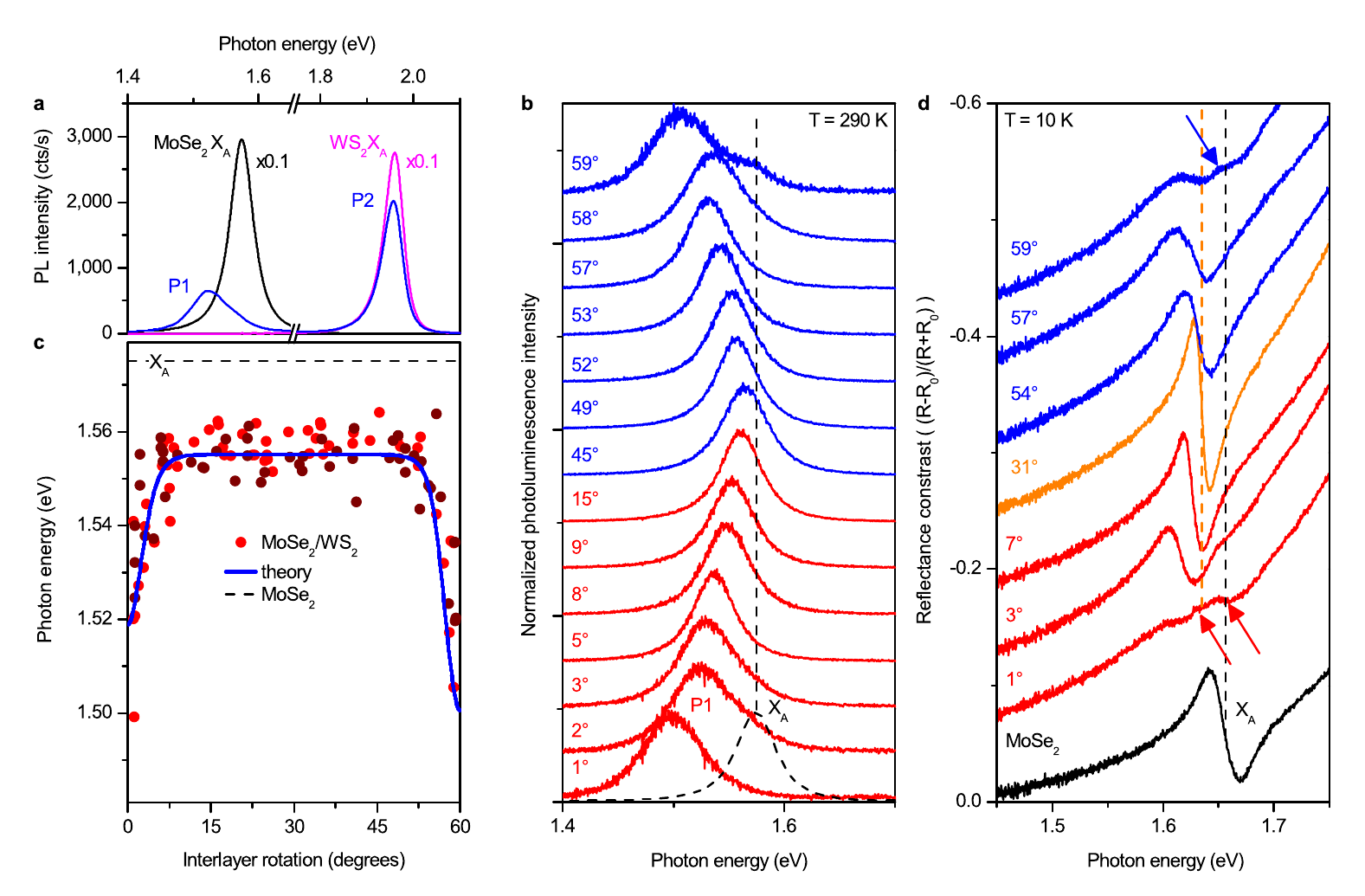}
	\caption{}
	\end{figure}
	
	\newpage

\begin{figure}[h!]
	\centering
	\includegraphics[width=1\columnwidth]{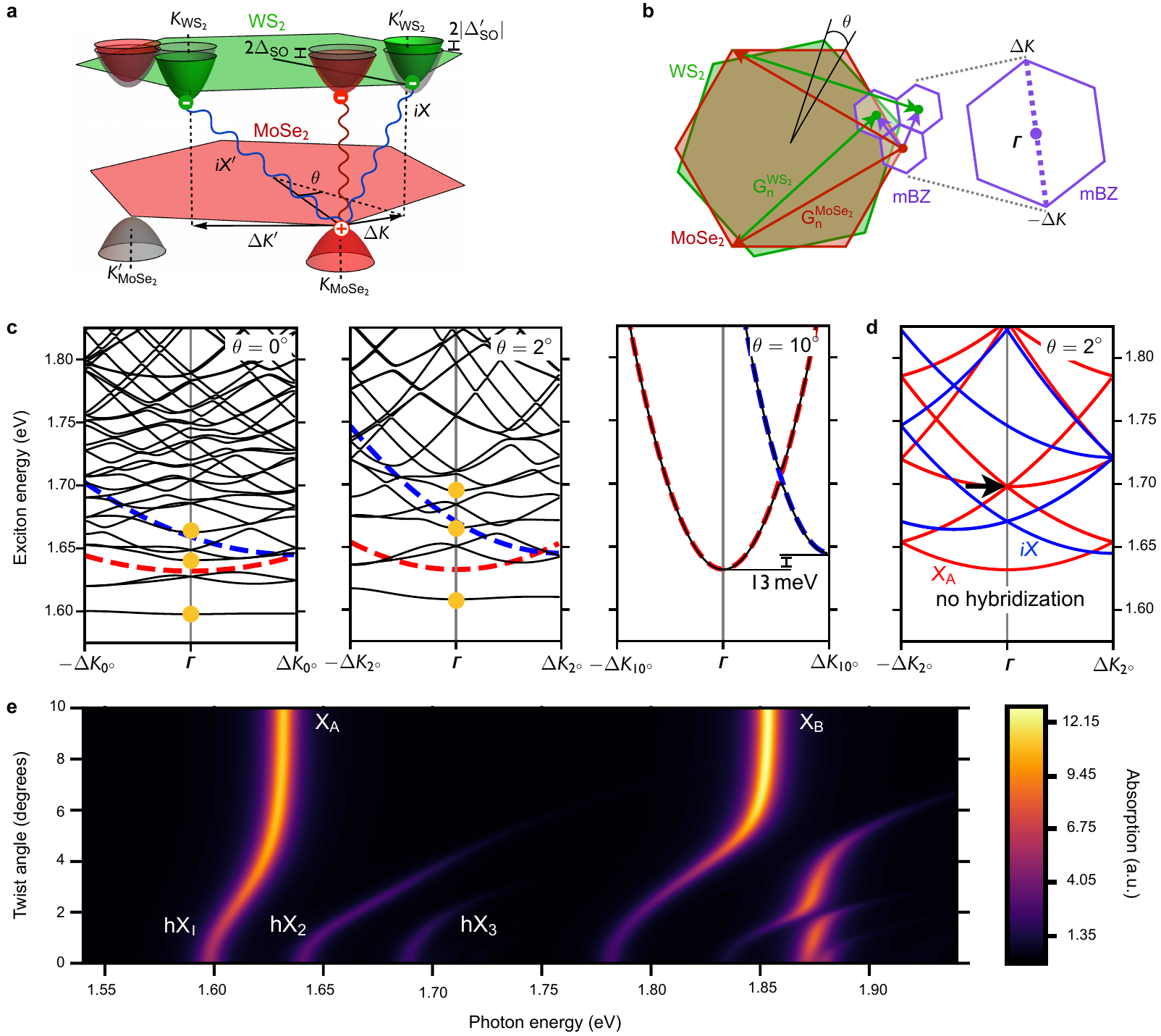}
	\caption{}
	\end{figure}
	
	\newpage

\begin{figure}[h!]
	\centering
	\includegraphics[width=0.8\columnwidth]{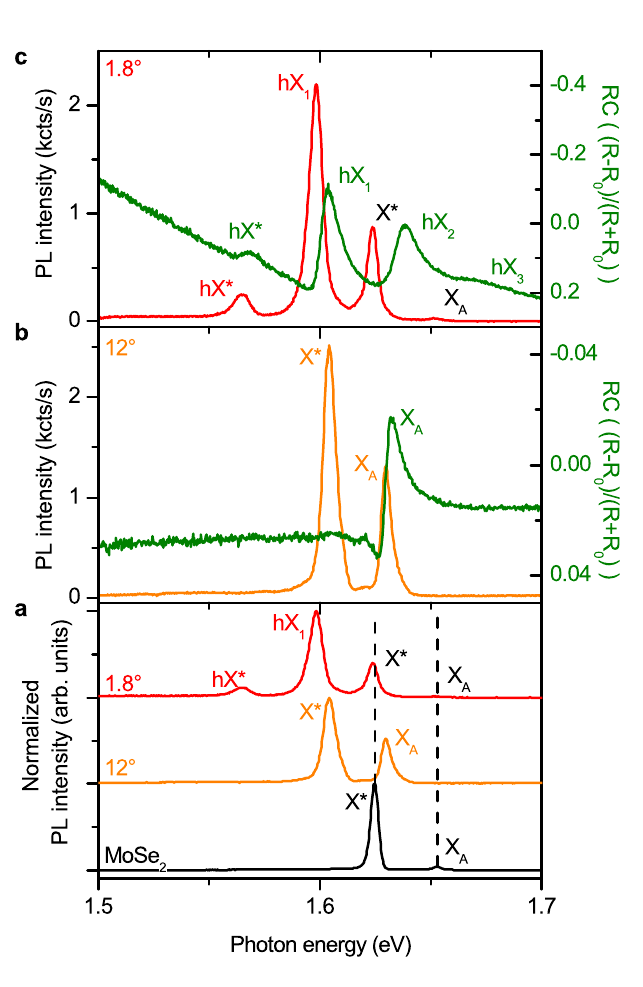}
	\caption{}
	\end{figure}
	
	\newpage

\begin{figure}[h!]
	\centering
	\includegraphics[width=0.55\columnwidth]{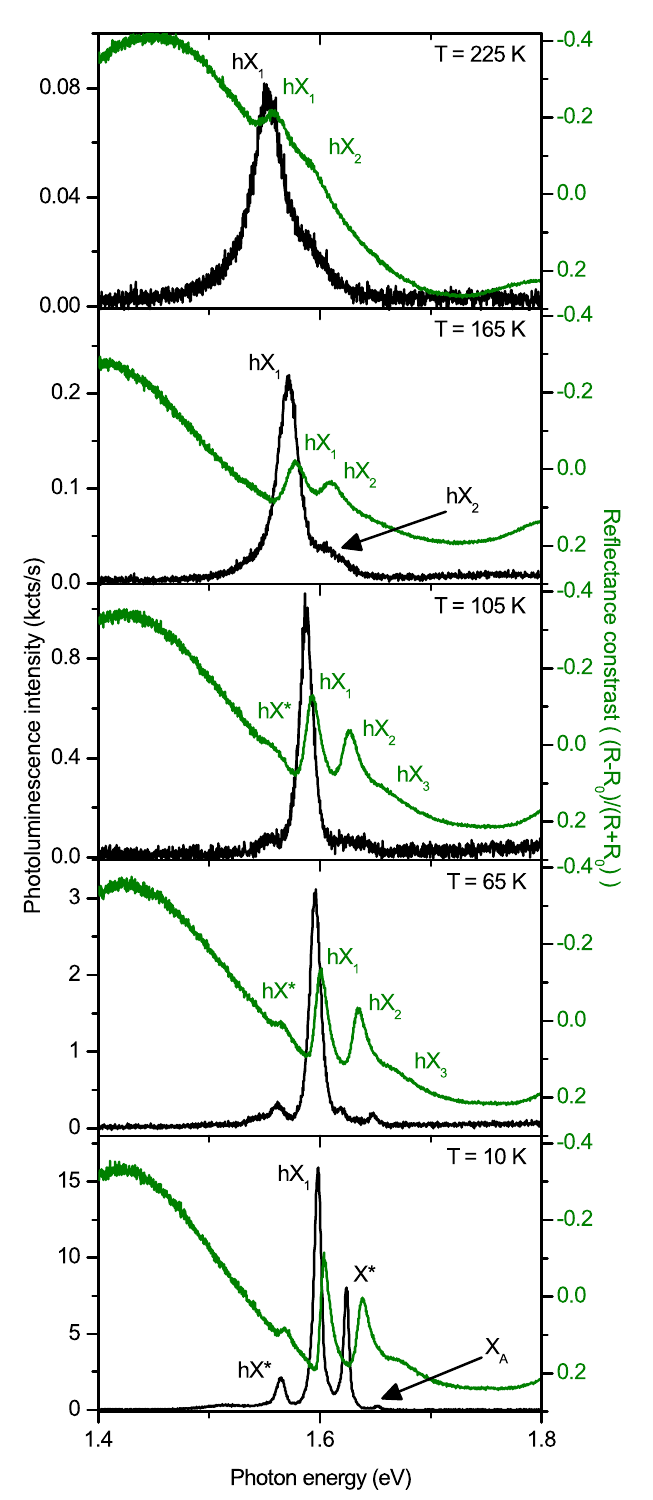}
	\caption{}
	\end{figure}

\newpage

\setcounter{figure}{0} 
\renewcommand{\figurename}{\textbf{Extended Data Fig.}}

\setcounter{table}{0} 
\renewcommand{\tablename}{\textbf{Extended Data Table}}

\newpage

\section{Extended Data Figure Captions and Extended Data Table}

\textbf{Extended Data Figure 1. Integrated intensity and linewidth for room-temperature PL spectra for MoSe$_2$/WS$_2$ heterobilayers as a function of the interlayer twist angle $\theta$.} (a) and (b) shows the variation of integrated intensity and linewidth, respectively.  See   {Methods} for the description of the fitting procedure. Data acquired on two individual substrates containing MoSe$_2$/WS$_2$ heterobilayers made from CVD-grown monolayers of MoSe$_2$ and WS$_2$ are shown with light and dark red symbols. The blue curve shows results of theoretical calculations as described in   {Supplementary Note~3}.\\
\label{fig:FWHM}

{\bf Extended Data Figure 2. Breakdown of the moir\'e harmonic potential approximation in MoSe${}_2$/WS${}_2$.} (a) MoSe${}_2$ A-exciton (X${}_{\rm A}$, red) and interlayer exciton (iX, blue) bands in MoSe${}_2$/WS${}_2$ within the moir\'e Brillouin zone, for twist angle $\theta = 54^\circ$.  Purple arrows represent second-order virtual tunnelling process enabled by intralayer-interlayer exciton hybridization, giving rise to a moir\'e potential for bright ($\Gamma$-point) MoSe${}_2$ A excitons. (b) Perturbation theory breaks down for MoSe${}_2$ $\Gamma$-point A excitons at $\theta = 58^\circ$, due to the exciton band crossing. As a result, the effective moir\'e potential diverges for this twist angle, indicating that the harmonic potential approximation is invalid. (c) Perturbative parameter $|\tilde{T}|/|\mathcal{E}_{\rm Y,\downarrow}^{+}(-\Delta\KK)-\mathcal{E}_{\rm X,\downarrow}^{+}(0)|$ (top, see   {Supplementary Note~2}) and P1 PL line energy (bottom). The vertical gray lines divide the plot into the $8^\circ \le \theta \le 52^\circ$ plateau and the two modulation regions, showing that the latter is always outside the region of validity of the harmonic approximation.\\
\label{fig:moire_breakdown}

{\bf Extended Data Figure 3. The moir\'e mini Brillouin zone in MoSe${}_2$/WS${}_2$ heterobilayers.} (a) Real-space stacking of the MoSe${}_2$ (red) and WS${}_2$ (green) lattices for twist angle $\theta$, with lattice vectors labeled $a_n^{\rm MX_2}$. Red (green) circles represent Mo (W) atoms, whereas Se and S atoms are shown as orange circles. (b) The resulting alignment between the Brillouin zones. Corresponding reciprocal lattice vectors $G_n^{\rm MoSe_2}$ and $G_n^{\rm WS_2}$ appear misaligned by the twist angle $\theta$. The monolayer band edges appear at the points $K_{\rm MX_2}$ of the corresponding BZs; the valley mismatch vectors are labeled $\Delta \KK$ and $\Delta \KK'$. (c) The two MoSe${}_2$ Bragg vectors that contribute to the hopping term in Eq.\ 3 in Supplementary Note 1. (d) First circle of Bragg vectors for the two TMD crystals, defining the moir\'e vectors $\bb_n = \GG_n^{\rm WS_2}-\GG_n^{\rm MoSe_2}$ for $\theta \approx 0^\circ$. (e) Mini Brillouin zone defined by the vectors $\bb_n$, where the lattice mismatch $\Delta\KK$ appears at the mBZ edge. The different moir\'e vectors can be constructed as $(\mathcal{D}-\mathcal{D}')\Delta\KK$. An example is shown, where $\Delta\KK-C_3\Delta\KK=-\bb_3$.\\
	
{\bf Extended Data Figure 4. Calculated moir\'e band structures of hybridized excitons in MoSe${}_2$/WS${}_2$ heterobilayers for various interlayer twist angles.} Bright hybridized exciton (hX) band structures within the first moir\'e Brillouin zone for various twist angles, calculated using the exfoliated sample parameters reported in   {Extended Data Table~I}. The red dashed curves in all panels show the uncoupled MoSe${}_2$ A exciton dispersion, for reference. In the top panels, the blue dashed curves corresponds to the iX interlayer exciton state described in the main text, whereas in the bottom panels they represent the interlayer exciton state labeled iX'.\\
\label{fig:moire_bands}
		
\textbf{Extended Data Figure 5. Low-temperature RC spectra of CVD-grown MoSe$_2$/WS$_2$ heterobilayers.} (a) RC spectra recorded at T = 10~K in CVD MoSe$_2$/WS$_2$ heterostructure with interlayer twist angle of 1${}^\circ$ (red), 31${}^\circ$ (orange) and 59${}^\circ$ (blue) in a spectral range around the MoSe$_2$ A exciton energy. Closely-aligned regions show the red shift and significant intensity reduction of the main peak compared to the rotationally misaligned heterobilayer, as well as the emergence of additional weak features above the main peak. (b) RC spectra measured in the vicinity of the B exciton energy in an isolated MoSe$_2$ monolayer (black), and in MoSe$_2$/WS$_2$ heterostructures for various $\theta$. The B exciton feature in the isolated MoSe$_2$ monolayer spectrum is labeled X${}_{\rm B}$. Vertical lines show the position of the maximum derivative of the X${}_{\rm B}$ feature in the monolayer (black) and misaligned heterobilayer (orange) spectra. See full description of the experimental procedure for the RC measurements in Methods.\\
\label{fig:NewLowTDataCVD}

\textbf{Extended Data Figure 6. Variation of PL and RC spectra in a MoSe$_2$/WS$_2$ heterobilayers made from exfoliated MoSe$_2$ and WS$_2$ monolayers encapsulated in hBN.} (a) Bright-field image of a fully-encapsulated MoSe$_2$/WS$_2$ sample S1, where the points for which we report the PL and RC spectra in (b)-(d) are marked. Scale bar corresponds to 10~$\mu$m.(b-d) Low-temperature PL (black) and RC (red) spectra recorded in several regions of the sample marked in (a). The two higher-energy peaks in the PL spectra, X* at 1.624 and X${}_{\rm A}$ at 1.652~eV, correspond to trion and excition emission unintentionally collected from the single-layer MoSe$_2$ area located at the right side of the heterobilayer sample. The position of these peaks remains unchanged in all three points, while their intensity decreases gradually with the increasing spatial separation. The two lower-energy PL peaks, labeled hX${}^*$ and hX${}_1$, represent the emission originating in the heterostructure region and show a variation of their position and relative intensities across the heterostructure region, likely caused by the non-uniform strain and doping. The RC spectra recorded in the three points are similar, with the two lower-energy peaks directly corresponding to hX${}^*$ and hX${}_1$ in PL, and hX${}_2$ and hX${}_3$ representing the higher-energy states. (e) Comparison of low-temperature PL spectra recorded in the samples fabricated from mechanically exfoliated monolayers. Dashed lines show PL spectra of uncoupled single-layer MoSe${}_2$, recorded in the same sample, where an uncoupled MoSe$_2$ monolayer area was present. Samples S1-S4 were fabricated with the crystal axes of the two materials closely aligned, whereas sample S5 was made with a significant rotational misalignment ($\theta = 12^\circ$). Despite the variation of exciton (X${}_{\rm A}$) and trion (X*) energies, all four aligned samples show a hybridised exciton peak hX${}_1$, located 20-30~meV below the monolayer trion line. Samples S1 and S2 show an additional lower-energy line hX${}^*$ positioned $\sim$32 meV below hX${}_1$ energy.   {Fig.~4} in the main text reports data for the closely-aligned sample S1 and the misaligned sample S5.   {Fig.~5} in the main text reports data for sample S1.\\

\textbf{Extended Data Figure 7. Temperature dependence of PL and RC spectra in a MoSe$_2$/WS$_2$ heterobilayer made from exfoliated MoSe$_2$ and WS$_2$ monolayers encapsulated in hBN.} The data presented here are for sample S1, for which additional data are presented in   {Figs.~4 and 5} in the main text. (a) Normalized PL spectra for hBN-encapsulated MoSe$_2$/WS$_2$ heterostructure S1 at different temperatures. At T=10 K, the emission spectrum consists of the hX${}^*$ peak at 1.565~eV and hX$_{1}$ peak 1.598~eV originating in the heterostructure, and the strong trion peak (X*) at 1.624 and a weaker neutral exciton peak (X${}_{\rm A}$) at 1.652~eV collected from the isolated MoSe$_2$ monolayer. hX${}^*$ disappears at $T \ge 105 \mathrm{K}$, while hX$_{1}$ is the dominant PL feature visible at room temperature. (b) RC spectra recorded in the same region of the sample and the same temperatures as the PL spectra in (a). The two lower-energy peaks visible in the low-temperature spectrum directly correspond to hX${}^*$ and hX${}_1$ PL features, whereas the peaks hX${}_2$ and hX${}_3$ represent the  higher-energy features which are not visible in PL. The hX${}^*$ and hX$_{3}$ peaks become weak above $T = 105 K$, while the hX$_{1}$ and hX$_{2}$ peaks persist up to much higher temperature, with the former remaining visible at room temperature. (c) Energy of hX${}^*$ (red), hX$_{1}$ (green), and hX$_{2}$ (blue) features in PL and RC as a function of temperature. The peak positions in PL (RC) are marked with a circle (triangle). (d) and (e) PL linewidth and integrated intensity of hX${}^*$ (red) and hX$_{1}$ (green) as functions of temperature. \\
		\label{fig:Tdependence}
	
{\bf Extended Data Figure 8. Theoretical PL spectra of MoSe${}_2$/WS${}_2$ for different temperatures.} (a) Calculated activation energy for hX${}_1$ PL, as a function of twist angle. (b) Normalised PL intensity in the MoSe${}_2$ A-exciton energy range, for three different temperatures: $T=60\,{\rm K},\,160\,{\rm K}$ and $300\,{\rm K}$ (room temperature). PL from state hX${}_1$ produces the peak identified as P1 in our CVD sample measurements. A second peak at higher photon energies (black arrow) is thermally activated at approximately $160\,{\rm K}$, corresponding to PL from the hX${}_2$ state, in excellent agreement with our exfoliated sample measurements   {(Fig.~5)}. PL peak broadening and red shift with increasing temperature are not taken into account for these simulations.\\
\label{fig:intensity}

\textbf{Extended Data Figure 9. Twist angle measurements using second harmonic generation (SHG)}. The symbols show the data for the nearly aligned (a) and misaligned (b) samples fabricated from mechanically exfoliated monolayers. Solid lines in the graph represent the fitting of the data with $I_{SHG} \propto \sin(3\alpha+\phi)$. All SHG measurements were taken at room temperature using a custom-built system with a Toptica FemtoFErb $\sim90$~fs SAM mode-locked laser with a repetition rate of 80~MHz centred at 785~nm. For each measurement, the laser light had a typical incident power of 500~$\mu$W, was linearly polarized and focused to a spot size of $<$ 2 $\mu$m by a 50x objective lens (NA = 0.36). The SHG signal was separated from the reflected light using a beam splitter and a short-pass filter. A second linear polarizer (analyzer) was placed in the SHG signal path and aligned parallel to the excitation polarization. A motorized half-wave plate was located above the objective and rotated in order to obtain angle resolution. The SHG signal was then measured by a spectrometer and a liquid nitrogen cooled charge-coupled device.
\label{fig:SHG}

\begin{table}[h!]
\caption{Parameters of the exciton hybridisation model (Supplementary Note 1,   {Eqs.\ 4a and 4b}). The MoSe${}_2$ A- and B-exciton energies ($E_{\rm X_A}$ and $E_{\rm X_B}$) were obtained from our measurements of strongly misaligned CVD (for room temperature) and exfoliated (for low temperature) samples. An intralayer-interlayer exciton detuning $E_{\rm iX}-E_{\rm X} = 13\,{\rm meV}$ and interlayer hopping strength $t_{cc}=26\,{\rm meV}$ were estimated from the room temperature CVD twist-angle dependent PL data for angles close to $0^\circ$. We have assumed that these parameters persist in the exfoliated samples, yielding the values reported below, which we used in our calculations of   {Fig.\ 3c-e} in the main text. The conduction and valence band spin-orbit splittings ($\Delta_{\rm SO}^{\rm c}$ and $\Delta_{\rm SO}^{\rm v}$), lattice constants ($a_{\rm MX_2}$), and momentum matrix elements between the conduction and valence bands at the valley ($\gamma$) were obtained from Ref.[32] in the main text. The exciton Bohr radii $a_{\rm X_A}$ and $a_{\rm iX}$, necessary to determine the intralayer-interlayer exciton hybridisation as a function of twist angle, were obtained from a finite--elements method solution for the exciton wavefunctions, using a dielectric constant $\epsilon = 2.45$ corresponding to a heterostructure placed between an SiO${}_2$ slab and vacuum (see Ref. [33] in the main text).}
\begin{center}
%\begin{tabular}{ccccc}
\begin{tabular*}{0.8\columnwidth}{@{\extracolsep{\stretch{1}}}*{7}{cccccccc}@{}}
\hline\hline
$E_{\rm X_A}$ (eV) & $E_{\rm X_B}$ (eV) & $E_{\rm iX}$ (eV) & $t_{cc}$ (meV) & $a_{\rm X_A}\, (\Ams)$ & $a_{\rm iX}\, (\Ams)$ \\ \hline\hline % & $E_{\rm X}$ [meV] & $E_{\rm Y}$ [meV] \\ \hline \hline
1.555\,\text{(CVD)} & 1.756\,\text{(CVD)} & 1.568\,\text{(CVD)} & \, & \, & \, \\
1.632\,\text{(exfoliated)} & 1.854\,\text{(exfoliated)} & 1.645\,\text{(exfoliated)} & \raisebox{2.2ex}[0pt]{26} & \raisebox{2.2ex}[0pt]{$16.5$} & \raisebox{2.2ex}[0pt]{$19.5$} \\ % & 196${}^\star$ & 162${}^\star$ \\
\hline \hline
%\end{tabular}
\end{tabular*}

\vspace{0.1cm}
\begin{tabular*}{0.8\columnwidth}{@{\extracolsep{\stretch{1}}}*{7}{r|cccccc}@{}}
%\begin{tabular}{cccccc}
\hline\hline
\, & $a_{\mathrm{MX}_2}\,[\Ams]$ & $m_{e}$ [$m_0$] & $m_{h}$ [$m_0$] & $\Delta_{\rm SO}^{\rm c}$ [meV] & $\Delta_{\rm SO}^{\rm v}$ [meV] & $\gamma\,[{\rm eV\cdot\AA}]$ \\ \hline \hline
MoSe${}_2$ & $3.289$ & $0.38$ & $0.44$ & $8.0$ & $93.0$ & 2.20 \\
WS${}_2$ & $3.16$ & $0.27$ & $0.32$ & $-16.0$ & $214.5$ & 2.59 \\
%MoS${}_2$ & $3.157$ & $0.35$ & $0.43$ & $1.5$ & $74.0$ & 2.22 \\
\hline \hline
%\end{tabular}
\end{tabular*}
\end{center}
\label{tab:parameters}
\end{table}%

\newpage

\renewcommand{\thesubsection}{\arabic{subsection}}

\section{Supplementary Notes}

\subsection{Model for intralayer-interlayer exciton hybridization and moir\'e superlattice minibands for excitons}
Bright MoSe${}_2$ excitons are formed in an MoSe${}_2$/WS${}_2$ heterobilayer by $\tau\KK_{\mathrm{MoSe}_2}$ valley electrons and $-\tau\KK_{\rm MoSe_2}$ holes of opposite spins, where $\tau=\pm 1$ is the valley index. For $\tau=1$, the intralayer exciton state is
\begin{equation}\label{eq:Xstate}
	\ket{{\rm X}_{s}(\QQ)} = \frac{1}{\sqrt{S}}\sum_{\kkappa}\tilde{\varphi}(\kkappa)c_{{\rm M},s}^\dagger(\KK_{\mathrm{MoSe}_2}+\tfrac{m_e}{M_{\rm X}}\QQ+\kkappa)h_{{\rm M},-s}^\dagger(-\KK_{\mathrm{MoSe}_2}+\tfrac{m_h}{M_{\rm X}}\QQ-\kkappa)\ket{\Omega},
\end{equation}
where $c_{\mathrm{M},s}^\dagger(\KK_{\mathrm{MoSe}_2}+\kk)$ [$h_{\mathrm{M},s}^\dagger(\KK_{\mathrm{MoSe}_2}+\kk)$] creates an electron (hole) in the MoSe${}_2$ spin--$s$ conduction (valence) band, with wave vector $\kk$ near $\KK_{\mathrm{MoSe}_2}$ ($s=\downarrow$ gives the A exciton, X${}_{\rm A}$, while $s=\uparrow$ gives the B exciton, X${}_{\rm B}$. The same is true for WS${}_2$), and $\tilde{\varphi}(\kkappa)$ is the Fourier transform of the (ground-state) exciton relative-motion wavefunction. $m_e$ and $m_h$ are the electron and hole effective masses; $M_{\rm X}=m_e+m_h$, and $\ket{\Omega}$ is the heterobilayer ground state.

MoSe${}_2$ electrons tunnel into the WS${}_2$ conduction band through the ``hopping term''\cite{Wang2016b},
\begin{equation*}\label{eq:interlayer_hopping}
\begin{split}
	t = \sum_{s,\tau'=\pm 1}\sum_{m,n}\sum_{\kk,\kk'}&\delta_{(\KK_{\mathrm{MoSe}_2}+\kk)-(\tau'\KK_{\mathrm{WS}_2}+\kk'),\GG_n^{\rm WS_2}-\GG_m^{\rm MoSe_2}}\,t_{cc}(\GG_m^{\rm MoSe_2}+\KK_{\mathrm{MoSe}_2}+\kk)\\
	&\times\left[\exp{-i\GG_m^{\rm MoSe_2}\cdot\rr_0}c_{\mathrm{W},s}^\dagger(\tau'\KK_{\mathrm{WS}_2}+\kk')c_{\mathrm{M},s}(\KK_{\mathrm{MoSe}_2}+\kk) + \mathrm{H.c.}\right],
\end{split}
\end{equation*}
where $\GG_m^{\rm MoSe_2}$ and $\GG_n^{\rm WS_2}$ are reciprocal lattice vectors of the corresponding crystals, and $\rr_0$ represents the in-plane shift between metal atoms in the two layers, which together with the twist angle $\theta$ parametrizes the heterobilayer stacking. The Kronecker delta encodes momentum conservation . Due to symmetry under $C_3$ rotations, the valley $-\KK_{\rm WS_2}$ is equivalent to $\KK_{\rm WS_2}'$ (see main text   {Fig.~3a}).

Intralayer MoSe${}_2$ excitons can hybridize with interlayer excitons (iXs) of same quantum number $s$, (WS${}_2$ electron and an MoSe${}_2$ hole)
\begin{equation}\label{eq:Ystate}
	|{\rm Y}^{\tau'}_{s}(\QQ')\rangle = \frac{1}{\sqrt{S}}\sum_{\kkappa}\tilde{\psi}(\kkappa)c_{W,s}^\dagger(\tau'\KK_{\mathrm{WS}_2}+\tfrac{m_e'}{M_{\rm iX}}\QQ'+\kkappa)h_{M,-s}^\dagger(-\KK_{\mathrm{MoSe}_2}+\tfrac{m_h}{M_{\rm Y}}\QQ'-\kkappa)\ket{\Omega},
\end{equation}
where $m_e'$ is the WS${}_2$ electron effective mass and $M_{\rm iX}=m_e'+m_h$ (see Extended Data Table I).

The relative-motion momentum-space wavefunctions of both exciton species are given by
\begin{equation*}
	\tilde{\varphi}(\kkappa) = \int\ud^2\rho\,\exp{-i\kkappa\cdot\rrho}\varphi(\rrho),\quad \tilde{\psi}(\kkappa) = \int\ud^2\rho\,\exp{-i\kkappa\cdot\rrho}\psi(\rrho),
\end{equation*}
and we obtained the real--space wavefunctions
\begin{equation*}\label{eq:sstate}
	\varphi(\rho) \approx \sqrt{\frac{2}{\pi a_{\rm X}^2}}\,\exp{-\rho/a_{\rm X}},\quad\psi(\rho) \approx \sqrt{\frac{2}{\pi a_{\rm iX}^2}}\,\exp{-\rho/a_{\rm iX}},
\end{equation*}
by solving numerically the two--body problem with bilayer Keldysh--type interactions \cite{keldysh,marcin_bindingenergies_prb_2017,mostaani_excitonic_prb_2017,complexes2018,berkelbach_prb_2013}, finding $a_{\rm X}$ and $a_{\rm iX}$ from the solutions. Then, we obtain the bright inter--intra exciton mixing term
\begin{equation}\label{eq:hopping_term}
\begin{split}
	T =& \sum_{s,\tau'}\sum_{m,n}\sum_{\QQ,\QQ'}T_{\tau'}(\GG_m^{\rm MoSe_2},\GG_n^{\rm WS_2})\delta_{\QQ-\QQ',\Delta\KK_{\tau'}+\GG_n^{\rm WS_2}-\GG_m^{\rm MoSe_2}}Y_{s}^{\tau'}{}^\dagger(\QQ')X_s(\QQ) + \mathrm{H.c.}\\
	T_{\tau'}(\GG,\GG') %=& t_{cc}(\KK_{\rm MoSe_2}+\GG)\exp{-i\GG\cdot\rr_0}\int\ud^2\rho\,\exp{-i\left[\tfrac{m_e}{M_{\rm X}}-\tfrac{m_e'}{M_{\rm Y}} \right]\rrho\cdot\QQ}\exp{-i\tfrac{m_h}{M_{\rm Y}}\rrho\cdot[\Delta\KK_{\tau'}+(\GG'-\GG)]}\varphi^*(\rho)\psi(\rho)\\
	\approx& \frac{4 t_{cc}(\KK_{\rm MoSe_2}+\GG)\exp{-i\GG\cdot\rr_0}}{a_{\rm X}a_{\rm iX}}\left(\frac{a_{\rm X}+a_{\rm iX}}{a_{\rm X}a_{\rm iX}}\right)%\\
	\left[\left(\frac{a_{\rm X}+a_{\rm iX}}{a_{\rm X}a_{\rm iX}} \right)^2 + \frac{m_h^2}{M_{\rm iX}^2}(\Delta \KK_{\tau'}+\GG'-\GG)^2 \right]^{-3/2},
\end{split}
\end{equation}
where $\Delta\KK_{\tau'}=\tau'\KK_{\mathrm{WS}_2}-\KK_{\mathrm{MoSe}_2}$; $\Delta\KK \equiv \Delta\KK_{+}$ and $\Delta\KK' \equiv \Delta\KK_{-}$; and $X_s(\QQ)$, $Y_s^{\tau'}(\QQ')$ are exciton annihilation operators.

The coupling function $t_{cc}(\qq)$ decays rapidly with wave vector for $|\qq|>|\KK_{\text{MoSe}_2}|$ \cite{macdonald_pnas,Wang2016b}, which allows us to set it as a constant $t_{cc}$ for $|\qq| \lesssim |\KK_{\text{MoSe}_2}|$, and zero otherwise. This makes $T_{\tau'}(\GG,\GG')$ finite only for $\GG=0$ and the two other MoSe${}_2$ Bragg vectors shown in   {Extended Data Fig.~3c}. For closely aligned ($\theta\approx 0^\circ$) configurations, when $\GG = \GG_n^{\rm MoSe_2}$, the hopping term gives significant contributions only if $\tau'=1$ and $\GG' = \GG_n^{\rm WS_2}$, and is vanishingly small otherwise. Thus, the allowed Bragg vector combinations give
\begin{equation*}
	\Delta\KK + \GG'-\GG =\left\{\begin{array}{ccc}
	\Delta\KK & , & \GG'=\GG=0\\
	C_3\Delta\KK & , & \GG'=\GG_2^{\rm WS_2},\,\GG=\GG_2^{\rm MoSe_2}\\
	C_3^2\Delta\KK & , & \GG'=-\GG_1^{\rm WS_2},\,\GG=-\GG_1^{\rm MoSe_2}
	\end{array}\right. .
\end{equation*}
Therefore, to a good approximation,
\begin{equation*}\label{eq:Top}
\begin{split}
	T =& \sum_{s}\sum_{\mathcal{D}}\sum_{\QQ,\QQ'}\delta_{\QQ-\QQ',\mathcal{D}\Delta\KK}\,T_{\mathcal{D}}\,Y_{s}^{+}{}^\dagger(\QQ')X_s(\QQ) + \mathrm{H.c.},\\
	T_{\mathcal{D}} =& \exp{i\KK\cdot\rr_0}\frac{4 t_{cc}\exp{-i\mathcal{D}\KK\cdot\rr_0}}{a_{\rm X}a_{\rm iX}}\left(\frac{a_{\rm X}+a_{\rm iX}}{a_{\rm X}a_{\rm iX}}\right)\left[\left(\frac{a_{\rm X}+a_{\rm iX}}{a_{\rm X}a_{\rm iX}} \right)^2 + \frac{m_h^2}{M_{\rm iX}^2}\Delta K^2 \right]^{-3/2},
\end{split}
\end{equation*}
where we define $\mathcal{D}\in\{E,\,C_3,\,C_3^2\}$, with $C_3^n$ a rotation by $\tfrac{2n\pi}{3}$ and $E$ the identity.

For $\theta \approx 60^\circ$ one must choose $\tau'=-1$ and for each Bragg vector $\GG=\GG_n^{\rm MoSe_2}$ take $\GG' = -C_3\GG_n^{\rm WS_2}$, resulting in
\begin{equation*}
	\tilde{T} = \sum_{s}\sum_{\mathcal{D}}\sum_{\QQ,\QQ'}\delta_{\QQ-\QQ',\mathcal{D}\Delta\KK'}\,\tilde{T}_{\mathcal{D}}\,Y_{s}^{-}{}^\dagger(\QQ')X_s(\QQ) + \mathrm{H.c.}
\end{equation*}

From the above analysis, we get for $\tau=1$ the exciton Hamiltonians
\begin{subequations}
\begin{equation}\label{eq:Hmoire0}
	H = \sum_{s}\sum_{\QQ}\left[\mathcal{E}_{\rm X,s}(\QQ)X_{s}^\dagger(\QQ)X_{s}(\QQ) + \mathcal{E}_{\rm iX,s}^+(\QQ)Y_{s}^{+}{}^{\dagger}(\QQ)Y_{s}^{+}(\QQ)\right]+T; \quad \theta <30^\circ,
\end{equation}
\begin{equation}\label{eq:Hmoire60}
	H = \sum_{s}\sum_{\QQ}\left[\mathcal{E}_{\rm X,s}(\QQ)X_{s}^\dagger(\QQ)X_{s}(\QQ) + \mathcal{E}_{\rm iX,s}^-(\QQ)Y_{s}^{-}{}^{\dagger}(\QQ)Y_{s}^{-}(\QQ)\right]+\tilde{T}; \quad \theta >30^\circ,
\end{equation}
\end{subequations}
where 
\begin{subequations}
\begin{equation*}
	\mathcal{E}_{\rm X,s}(\QQ) = E_{\rm X}^0 + s (\Delta_{\rm SO}^{\rm v} + \Delta_{\rm SO}^{\rm c}) + \frac{\hbar^2Q^2}{2M_{\rm X}},
\end{equation*}
\begin{equation*}
	\mathcal{E}_{\rm iX,s}^{\tau'}(\QQ) = E_{\rm iX}^0 + s (\Delta_{\rm SO}^{\rm v}+\tau'\Delta_{\rm SO}^{\rm c}{}') + \frac{\hbar^2Q^2}{2M_{\rm iX}},
\end{equation*}
\end{subequations}
with $\Delta_{\mathrm{SO}}^{\rm c}$ and $\Delta_{\mathrm{SO}}^{\rm v}$ the spin-orbit couplings of the MoSe${}_2$ conduction and valence bands, and $\Delta_{\mathrm{SO}}^{\rm c}{}'$ the WS${}_2$ conduction band spin-orbit coupling. Thus, for $Q=0$, the A- and B-exciton energies can be written as (see   {Extended Data Table I}) $E_{\rm X_A}=E_{\rm X}^0 - (\Delta_{\rm SO}^{\rm v} + \Delta_{\rm SO}^{\rm c})$ and $E_{\rm X_B}=E_{\rm X}^0 + (\Delta_{\rm SO}^{\rm v} + \Delta_{\rm SO}^{\rm c})$. Analogous terms exist for the $\tau=-1$ valley, given by a time reversal transformation.

The moir\'e superlattice periodicity, introduced in Eqs.\ \eqref{eq:Hmoire0} and \eqref{eq:Hmoire60} through the terms $T$ and $\tilde{T}$, requires that we fold the X and iX bands onto the moir\'e Brillouin zone,
\begin{subequations}
\begin{equation*}
    \ket{{\rm X}_{s}(\QQ)}_{m,n}\equiv \ket{X_s(\QQ+m\bb_1+n\bb_2)},
\end{equation*}
\begin{equation*}
    |{\rm Y}_{s}^{\tau'}(\QQ')\rangle_{m,n}\equiv |Y_s^{\tau'}(\QQ'+m\bb_1+n\bb_2)\rangle,
\end{equation*}
\end{subequations}
where $\QQ$ is limited to the first moir\'e Brillouin zone (mBZ,   {Extended Data Fig.~3e}). The intra- and interlayer exciton states $| {\rm X}_s(\QQ) \rangle_{m,n}$ and $|{\rm Y}_s^{\tau'}(\QQ') \rangle_{m',n'}$ hybridize when $\QQ=\QQ'$ and
\begin{equation*}
    (m'-m)\bb_1 + (n'-n)\bb_2 = \bb_j\,;\,j=\pm1,\pm2,\pm3,
\end{equation*}
producing hXs states
\begin{equation*}
    | {\rm hX}_s^{\tau'}(\QQ) \rangle_{i,j} \equiv \sum_{m,n=0}^\infty\left[A_{i,j}^{m,n}(s,\QQ)|{\rm X}_s(\QQ) \rangle_{m,n} + B_{i,j}^{m,n}(s,\tau',\QQ)|{\rm Y}_s^{\tau'}(\QQ) \rangle_{m,n} \right],\, \QQ\in \text{mBZ},
\end{equation*}
with corresponding energies $E_{s;i,j}^{\tau'}(\QQ)$.

To evaluate the optical spectra of hX states, we use the light-matter interaction Hamiltonian
\begin{equation*}
    H_{\rm LM}=\frac{e\gamma}{\hbar c}\sum_{s}\sum_{\eta=\pm1}\sum_{\xxi,\xi_z}\sum_{\kk}\sqrt{\frac{4\pi\hbar c}{V\xxi}}c_{\rm M,s}^\dagger(\eta\KK_{\rm MoSe_2}+\kk-\xxi)h_{\rm M,s}^\dagger(-\eta\KK_{\rm MoSe_2}-\kk)a_{\eta}^\dagger(\xxi,\xi_z) + \mathrm{H.c.}
\end{equation*}
Here, $a_{\eta}^\dagger(\xxi,\xi_z)$ creates a photon of in-plane momentum $\xxi$ and out-of-plane momentum $\xi_z$, and polarization $\eta=\pm1$, corresponding to counter-clockwise and clockwise, respectively. We obtain the recombination and absorption rates from Fermi's golden rule ($\QQ + m\bb_1+n\bb_2 = \xxi$):
\begin{subequations}
\begin{equation*}\label{eq:PLeq}
    \Gamma_{\rm PL;m,n;s}^{\tau'}(\QQ)=\frac{4\pi}{\hbar}\sum_{\xxi,\xi_z}|\langle \eta;\xxi,\xi_z |H_{\rm LM} | {\rm hX}_s^{\tau'}(\QQ) \rangle_{m,n}|^2n_{\rm B}(E_{s;m,n}^{\tau',T})\delta\left( E_{s;m,n}^{\tau'}(\QQ) - \hbar c\sqrt{|\xxi|^2+\xi_z^2} \right),
\end{equation*}
\begin{equation*}\label{eq:ABSeq}
    \Gamma_{\rm A}^{\eta}(\xxi,\xi_z)=\frac{2\pi}{\hbar}\sum_s\sum_{m,n}\sum_{\QQ}|{}_{m,n}\langle  {\rm hX}_s^{\tau'}(\QQ) |H_{\rm LM} | \eta;\xxi,\xi_z \rangle|^2\delta\left( E_{s;m,n}^{\tau'}(\QQ) - \hbar c\sqrt{|\xxi|^2+\xi_z^2} \right),
\end{equation*}
\end{subequations}

For PL, we take into account temperature effects through the Bose-Einstein distribution
\begin{equation*}
    n_{\rm B}(E,T) = \frac{1}{\exp{(E-E_{\rm gnd})/k_{\rm B}T}+1},
\end{equation*}
where $E_{\rm gnd}$ is the energy of the lowest exciton state. The calculated twist-angle dependence of the activation energy $E_{\downarrow;0,0}(0)-E_{\rm gnd}$ for hX${}_1$ in MoSe${}_2$/WS${}_2$, as well as PL spectra at several temperatures, are shown in    {Extended Data Fig.~8}.

For absorption, we find \cite{hX_theory}
\begin{equation*}
    I_{\rm A}^s(\hbar \omega)=\frac{8\omega \delta\omega}{\hbar \pi c^2}\frac{e^2}{\hbar c}\sum_{m,n}\left|\sum_{i,j}\frac{\gamma A_{m,n}^{i,j}(s,0)}{a_{\rm X}} \right|^2\frac{\beta/\pi}{(\hbar \omega - E_{s;m,n}^{\tau'}(0))^2+\beta^2},
\end{equation*}
where we use $\beta = 5\,{\rm meV}$ and $\hbar\delta \omega=1\,{\rm meV}$ to evaluate the spectrum shown in   {Fig.~3e}.

\subsection{Harmonic potential approximation to exciton moir\'e effects in M\MakeLowercase{o}S\MakeLowercase{e}${}_2$/WS${}_2$ heterostructures}\label{sec:moire}
The moir\'e superlattice effects on the band structure \cite{wallbank,uchoa} and exciton energies \cite{macdonald_pnas} of bilayer systems, produced by incommensurability and misalignment of the two lattices, are often described in terms of a minimal harmonic potential \cite{mcdonald_intra,hongyi_moire,Wu2017}. In this section we derive the tunnelling contribution to this potential for intralayer excitons in TMD heterobilayers, using MoSe${}_2$/WS${}_2$ as a case study. We show that a harmonic potential fails to describe the moir\'e superlattice effects in the close alignment (anti-alignment) regime in the case of near-resonant exciton bands.

For $\theta < 30^\circ$, interlayer tunnelling $T$ allows MoSe${}_2$ intralayer excitons to explore the reciprocal lattice of the WS${}_2$ layer through virtual tunneling of their electrons onto the WS${}_2$ conduction band, and then back onto the MoSe${}_2$ conduction band. These virtual processes introduce momentum-dependent corrections to the intralayer exciton energies, which in real space correspond to a potential. 

Focusing on the case of $\theta < 30^\circ$, we perform a canonical transformation $\tilde{H} = \exp{iS}H\exp{-iS}$ on the intralayer-interlayer exciton Hamiltonian $H=H_0+T$ presented in main text Eq.\ (4a), with the condition \cite{schrieffer_wolff}
\begin{equation}\label{eq:SW}
T = -i\comm{S}{H_0},
\end{equation}
which removes from $\tilde{H}$ all terms that are first order in $T$. A similar procedure is followed for the Hamiltonian of main text Eq.\ (4b), for $\theta > 30^\circ$. The condition Eq.\ \eqref{eq:SW} is achieved by the generator 
\begin{equation}
iS = \sum_{s}\sum_{\mathcal{D}}\sum_{\QQ,\QQ'}\delta_{\QQ,\QQ'+\mathcal{D}\Delta\KK}\left[\frac{T_{\mathcal{D}}}{\mathcal{E}_{\rm iX,s}^{+}(\QQ')-\mathcal{E}_{\rm X,s}(\QQ)}Y_{s}^{+}{}^\dagger(\QQ')X_s(\QQ) - \mathrm{H.c.} \right]
\end{equation}
Evaluating $\tilde{H}$ up to second order in $T$ we obtain
\begin{equation}\label{eq:moire}
\begin{split}
&\tilde{H} \approx \sum_{s}\sum_{\QQ}\left[\tilde{E}_{\rm X,s}(\QQ)X_{s}^\dagger(\QQ)X_{s}(\QQ) + \tilde{E}_{\rm iX,s}(\QQ)Y_{s}^{+}{}^{\dagger}(\QQ)Y_{s}^{+}(\QQ)\right]\\
&+\frac{1}{2}\sum_{s}\sum_{\mathcal{D}\ne\mathcal{D}'}\sum_{\QQ}\left[\frac{T_{\mathcal{D}}^*T_{\mathcal{D}'}}{\mathcal{E}_{\rm iX,s}^{+}(\QQ)-\mathcal{E}_{\rm X,s}^+(\QQ+\mathcal{D}\Delta\KK)}Y_{s}^{+}{}^\dagger(\QQ+[\mathcal{D}-\mathcal{D}']\Delta\KK)Y_s^{+}(\QQ) + \mathrm{H.c.} \right]\\
&-\frac{1}{2}\sum_{s}\sum_{\mathcal{D}\ne\mathcal{D}'}\sum_{\QQ}\left[\frac{T_{\mathcal{D}}^*T_{\mathcal{D}'}}{\mathcal{E}_{\rm iX,s}^{+}(\QQ-\mathcal{D}\Delta\KK)-\mathcal{E}_{\rm X,s}^+(\QQ)}X_{s}^\dagger(\QQ+[\mathcal{D}-\mathcal{D}']\Delta\KK)X_s(\QQ) + \mathrm{H.c.} \right],
\end{split}
\end{equation}
with the renormalized energies
\begin{subequations}
	\begin{equation}
	\tilde{E}_{\rm X,s}^\tau(\QQ) = \mathcal{E}_{\rm X,s}^\tau(\QQ)-\frac{1}{2}\sum_{\mathcal{D}}\frac{|T_{\mathcal{D}}|^2}{\mathcal{E}_{\rm Y,s}^{\tau}(\QQ-\mathcal{D}\Delta\KK_{\tau})-\mathcal{E}_{\rm X,s}^{\tau}(\QQ)},
	\end{equation}
	\begin{equation}
	\tilde{E}_{\rm iX,s}^{\tau\tau'}(\QQ) = \mathcal{E}_{\rm Y,s}^{\tau}(\QQ)+\frac{1}{2}\sum_{\mathcal{D}}\frac{|T_{\mathcal{D}}|^2}{\mathcal{E}_{\rm Y,s}^{\tau}(\QQ)-\mathcal{E}_{\rm X,s}^{\tau}(\QQ+\mathcal{D}\Delta\KK_{\tau})}.
	\end{equation}
\end{subequations}
The remaining two terms represent scattering by moir\'e vectors $\bb = (\mathcal{D}-\mathcal{D}')\Delta\KK_{\tau}$, as shown in   {Extended Data Fig.~3e}. For exciton momenta near the center of the moir\'e Brillouin zone we have $Q \ll \Delta K$, and we may approximate
\begin{subequations}
	\begin{equation}
	\frac{T_{\mathcal{D}}^*T_{\mathcal{D}'}}{\mathcal{E}_{\rm iX,s}^{+}(\QQ)-\mathcal{E}_{\rm X,s}^{+}(\QQ+\mathcal{D}\Delta\KK)} \approx \frac{T_{\mathcal{D}}^*T_{\mathcal{D}'}}{[(E_{\rm iX}^0-\Delta_{\rm SO}')-(E_{\rm X}^0-\Delta_{\rm SO})] -\tfrac{\hbar^2\Delta K^2}{2M_{\rm X}} },
	\end{equation}
	\begin{equation}
	\frac{T_{\mathcal{D}}^*T_{\mathcal{D}'}}{\mathcal{E}_{\rm iX,s}^{+}(\QQ-\mathcal{D}\Delta\KK)-\mathcal{E}_{\rm X,s}^{+}(\QQ)} \approx \frac{T_{\mathcal{D}}^*T_{\mathcal{D}'}}{[(E_{\rm iX}^0-\Delta_{\rm SO}')-(E_{\rm X}^0 -\Delta_{\rm SO})] +\tfrac{\hbar^2\Delta K^2}{2M_{\rm iX}} }.
	\end{equation}
\end{subequations}
Finally, an inverse Fourier transform gives the harmonic potential for bright intralayer excitons at valley $\tau=1$
\begin{equation}\label{eq:harm_0}
V_{\rm X,s}(\rr) = \sum_{n=1}^3\frac{T_{C_3^{n-1}}^*T_{C_3^{n-2}}\exp{-i\dd_n\cdot\rr} + T_{C_3^{n-2}}^*T_{C_3^{n-1}}\exp{i\dd_n\cdot\rr} }{[(E_{\rm iX}^0-\Delta_{\rm SO}')-(E_{\rm X}^0 -\Delta_{\rm SO})] +\tfrac{\hbar^2\Delta K^2}{2M_{\rm iX}} },
\end{equation}
where $C_3^0 = E$, and for convenience we have re-labeled the moir\'e Bragg vectors as follows: $\dd_1 = \bb_1$, $\dd_2 = \bb_3$ and $\dd_3 = -\bb_2$. A similar analysis for $\theta > 30^\circ$ leads to the potential
\begin{equation}\label{eq:harm_60}
W_{\rm X,s}(\rr) = \sum_{n=1}^3\frac{\tilde{T}_{C_3^{n-1}}^*\tilde{T}_{C_3^{n-2}}\exp{-i\tilde{\dd}_n\cdot\rr} + \tilde{T}_{C_3^{n-2}}^*\tilde{T}_{C_3^{n-1}}\exp{i\tilde{\dd}_n\cdot\rr} }{[(E_{\rm iX}^0+\Delta_{\rm SO}')-(E_{\rm X}^0 -\Delta_{\rm SO})] +\tfrac{\hbar^2\Delta K'{}^2}{2M_{\rm iX}} }.
\end{equation}

Using the values of Extended Data Table I, we find that $W_{\rm X,s}(\rr)$ diverges at $\theta \approx 58^\circ$, signaling the breakdown of perturbation theory due to a crossing between the intralayer and interlayer exciton bands at the iX band edge, as shown in   {Extended Data Fig.\ 2b}. Furthermore,   {Extended Data Fig.\ 2c} shows that, although $V_{\rm X,s}(\rr)$ remains finite for all $\theta < 30^\circ$, the excitation energy in the virtual process becomes smaller than the mixing energy for $\theta < 5^\circ$, indicating that the perturbative approach is no longer valid.

Beyond these angles the intralayer-interlayer exciton mixing strength becomes the dominant energy scale in the problem, such that perturbative methods in general, and a simple description in terms of a potential in particular, cannot describe hX states or the moir\'e superlattice effects. This is a direct consequence of the near-resonant conduction bands in MoSe${}_2$/WS${}_2$ heterostructures.

\subsection{Broadening of the photoluminescence line by random fields in the sample}\label{sec:fwhm}

The dependence on twist angle of the emission line broadening shown in   {Extended Data Fig.\ 1b} may be explained by the coupling of weak electric fields produced by random strain throughout the sample, with the out-of-plane electric dipole of the mixed intralayer-interlayer exciton states. The field-dipole coupling can be estimated as
\begin{equation}
\begin{split}
	H_{\rm E-D} =& -\frac{ed\,E_z}{2}\sum_{s}\sum_{\tau=\pm1}\left[ \sum_{\kk} c_{{\rm M},s}^\dagger(\tau\KK_{\text{MoSe}_2}+\kk) c_{\rm M,s}(\tau\KK_{\text{MoSe}_2}+\kk) \right.\\
	& \qquad\qquad- \left. \sum_{\kk'} c_{{\rm W},s}^\dagger(\tau\KK_{\text{WS}_2}+\kk') c_{\rm W,s}(\tau\KK_{\text{WS}_2}+\kk') \right]\\
	&+\frac{ed\,E_z}{2}\sum_{s}\sum_{\tau=\pm1}\left[ \sum_{\kk} h_{{\rm M},s}^\dagger(\tau\KK_{\text{MoSe}_2}+\kk) h_{\rm M,s}(\tau\KK_{\text{MoSe}_2}+\kk) \right.\\
	& \qquad\qquad- \left. \sum_{\kk'} h_{{\rm W},s}^\dagger(\tau\KK_{\text{WS}_2}+\kk') h_{\rm W,s}(\tau\KK_{\text{WS}_2}+\kk') \right],
\end{split}
\end{equation}
where $c_{\rm M,s}(\qq)$ and $c_{\rm W,s}(\qq)$ annihilate an electron of momentum $\qq$ and spin projection $s$ in MoSe${}_2$ and WS${}_2$, respectively; $e$ is the charge unit, $d$ the interlayer distance, and we assume that the out-of-plane electric field $E_z$ is small. In first--order perturbation theory, this gives a correction to the bright, optically--active ($Q=0$) mixed exciton energy
\begin{equation}
	\delta E={}_{0,0}\braoket{{\rm hX}_{\downarrow}(0)}{H_{\rm E-D}}{{\rm hX}_{\downarrow}(0)}_{0,0}=ed\,E_z\,\abs{\langle{\rm Y}_{\downarrow}^{\tau'}(0)|{\rm hX}_{\downarrow}(0)\rangle_{0,0}}^2,
\end{equation}
where $\tau' = 1$ ($\tau'=-1$) for $\theta < 30^\circ$ ($\theta \ge 30^\circ$), $| {\rm hX_{\downarrow}(0)} \rangle_{0,0}$ is the lowest bright hybridized exciton state, and $\langle{\rm Y}_{\downarrow}^{\tau'}(0)|{\rm hX}_{\downarrow}(0)\rangle_{0,0}$ is its interlayer exciton component (see Methods in main text). Excitons in different parts of the sample will experience different values of $E_z$. Assuming that the $E_z$ values found throughout the sample follow a Gaussian distribution
\begin{equation}
	\rho(E_z) = \sqrt{\frac{1}{2\pi\sigma^2}}\,\exp{-(E_z - E_z^0)^2/2\sigma^2},
\end{equation}
of mean $E_z^0$ and variance $\sigma^2$, the correction $\delta E$ will also be normally distributed, with mean value
\begin{equation}
	\left< \delta E \right>_{\rho} = ed\,E_z^0\,\abs{\langle{\rm Y}_\downarrow^{\tau'}(0)|{\rm hX}_\downarrow (0)\rangle_{0,0}}^2,
\end{equation}
and a full width at half maximum given by
\begin{equation}\label{eq:FWHM}
	\text{FWHM}_{\delta E}=2\sqrt{2\log 2}\sqrt{\langle \delta E^2 \rangle_\rho - \langle \delta E \rangle_\rho^2} = 2\sigma\sqrt{2\log 2}\,ed\abs{\langle{\rm Y}_\downarrow^{\tau'}(0)|{\rm hX}_\downarrow(0)\rangle_{0,0}}^2.
\end{equation}
Allowing for an additive constant, representing the intrinsic broadening of the PL line, we fitted Eq.\ (\ref{eq:FWHM}) to the experimental data, and the result is presented in   { Extended Data Fig.\ 1b} of the main text. The fitting parameters give $\sigma e d = 19.8\,{\rm meV}$, or $\sigma \approx 0.03\,{\rm V}/{\rm nm}$, assuming an approximate interlayer distance of $6\,\Ams$ \cite{mattheiss_tmds_1973}.

\section{References}
%\bibliography{library}
\bibliographystyle{naturemag}

\newpage

\section{Extended Data Figures}

\begin{figure}[hbt!]
	\centering
	\includegraphics[width=1\columnwidth]{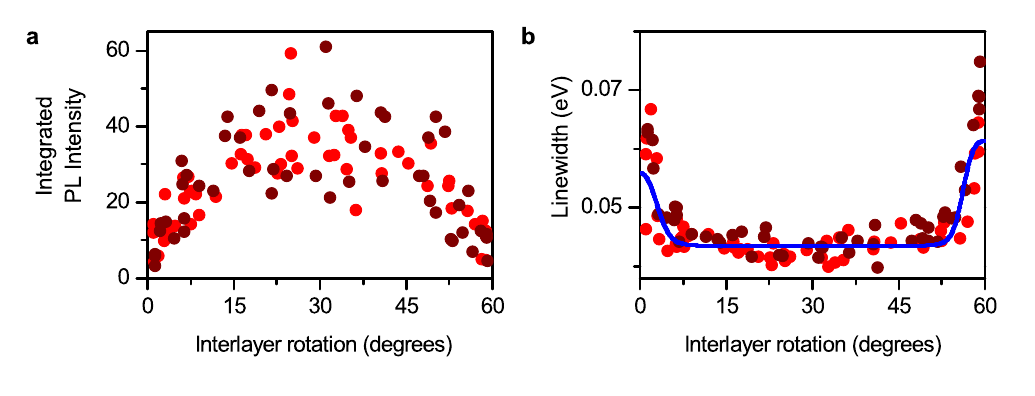}
	\caption{}
	\end{figure}
	
	%\newpage

\begin{figure}[hbt!]
	\centering
	\includegraphics[width=1\columnwidth]{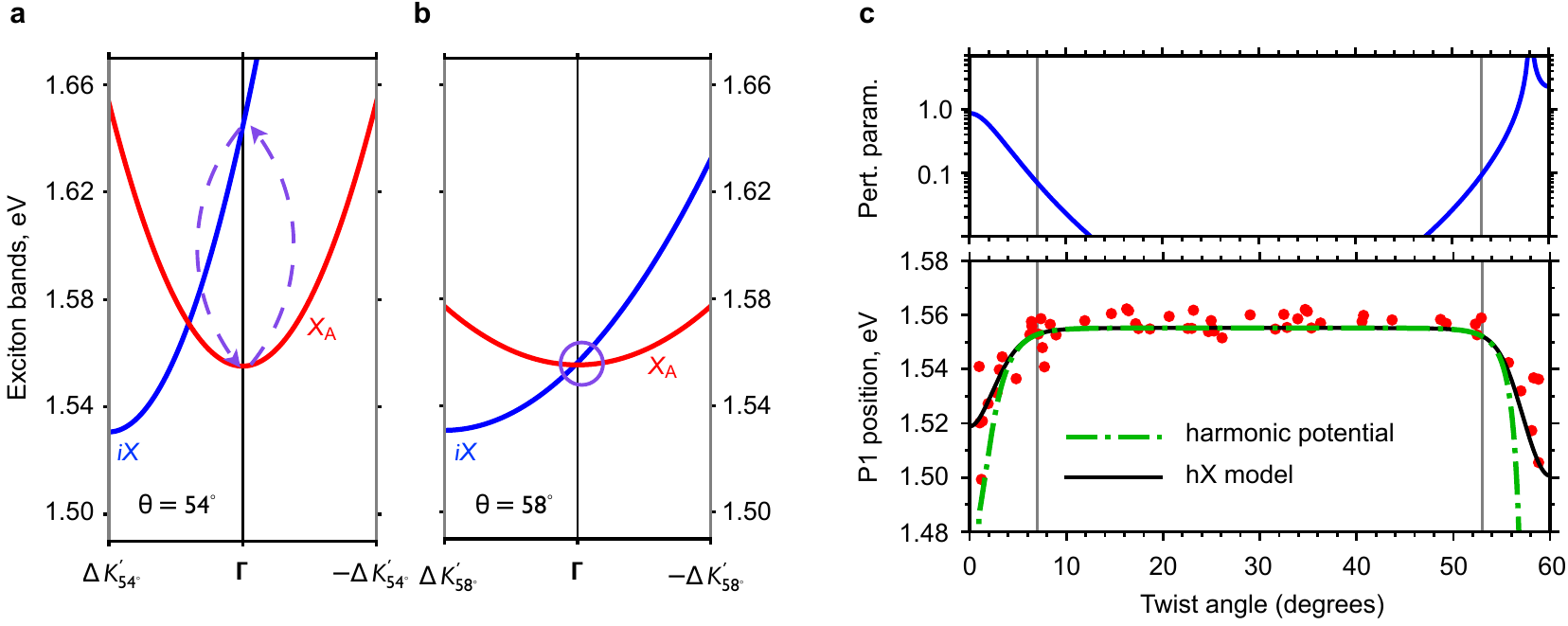}
	\caption{}
	\end{figure}
	
	%\newpage

\begin{figure}[hbt!]
	\centering
	\includegraphics[width=1\columnwidth]{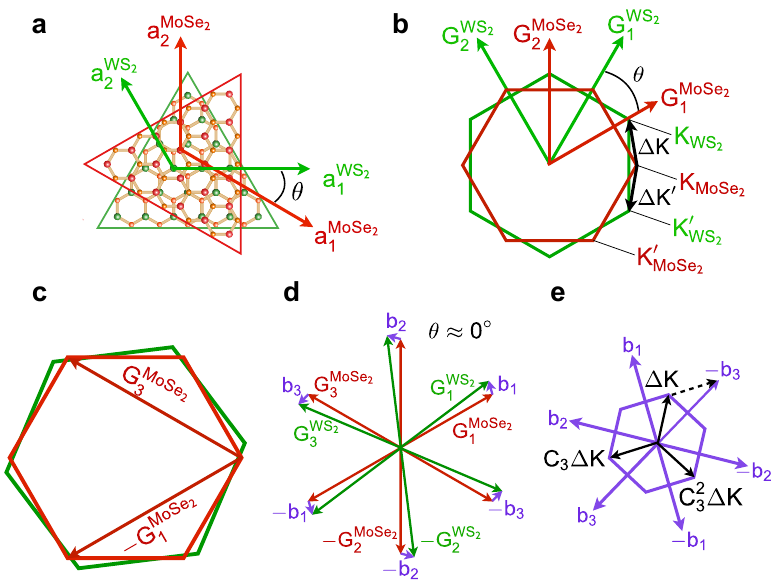}
	\caption{}
	\end{figure}
	
	%\newpage

\begin{figure}[hbt!]
	\centering
	\includegraphics[width=1\columnwidth]{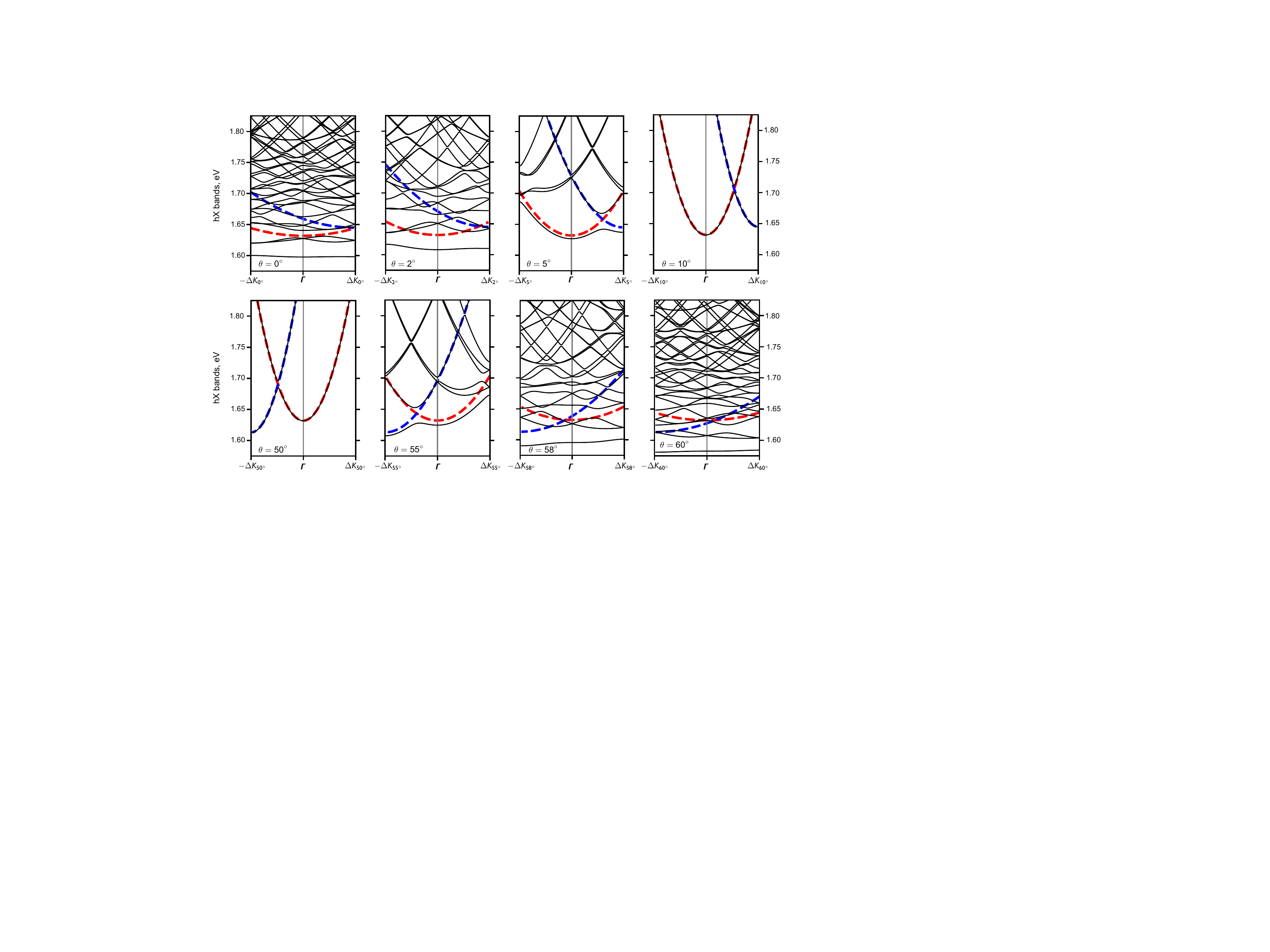}
	\caption{}
	\end{figure}
	
	%\newpage

\begin{figure}[hbt!]
	\centering
	\includegraphics[width=1\columnwidth]{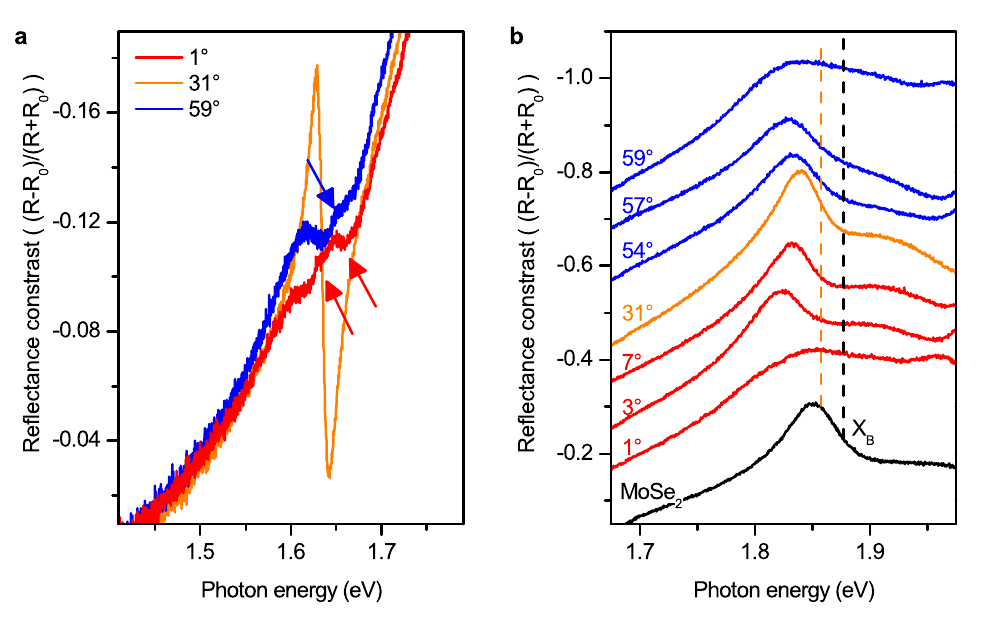}
	\caption{}
	\end{figure}
	
	%\newpage

\begin{figure}[hbt!]
	\centering
	\includegraphics[width=1\columnwidth]{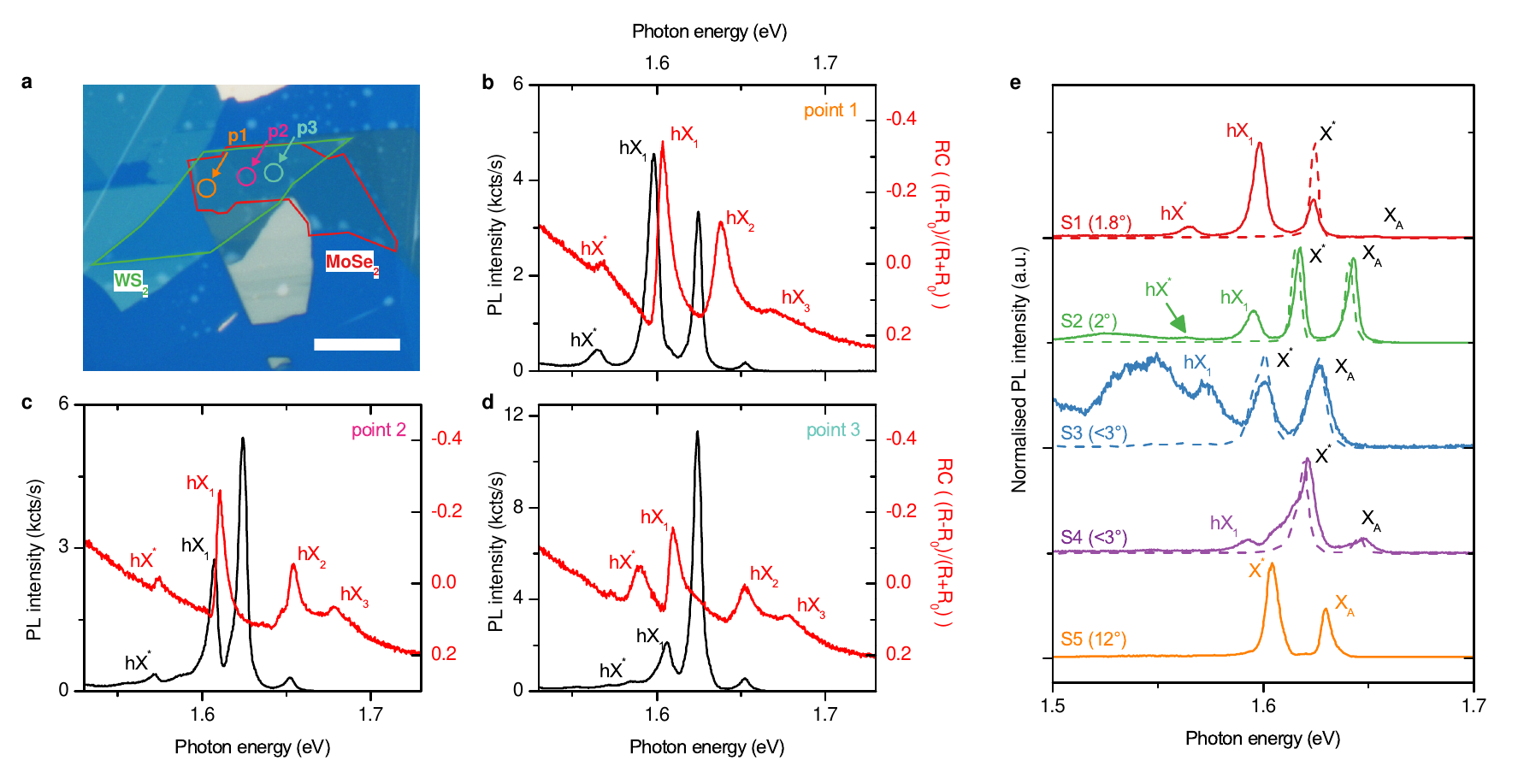}
	\caption{}
	\end{figure}
	
	%\newpage

\begin{figure}[hbt!]
	\centering
	\includegraphics[width=1\columnwidth]{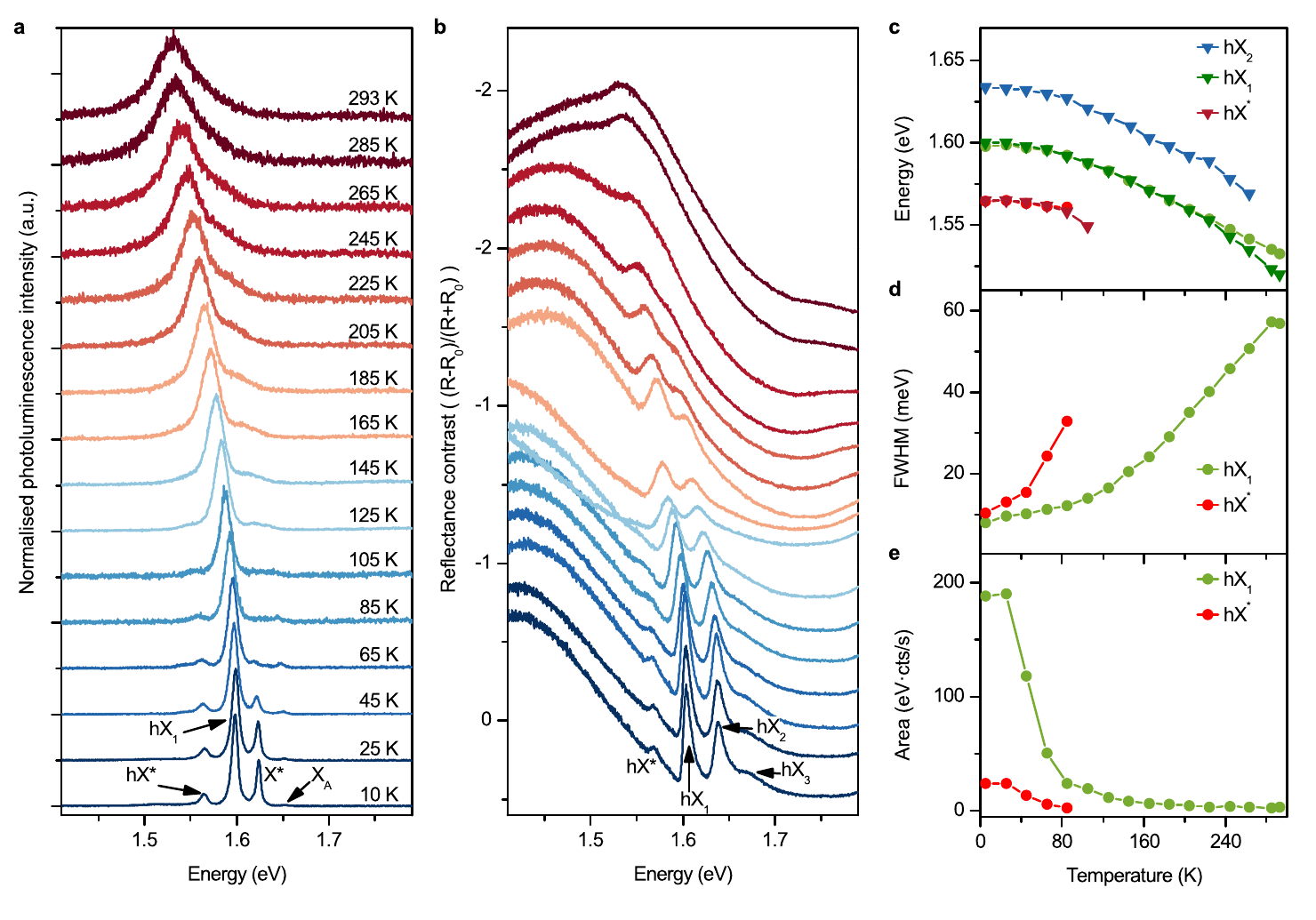}
	\caption{}
	\end{figure}
	
	%\newpage

\begin{figure}[hbt!]
	\centering
	\includegraphics[width=1\columnwidth]{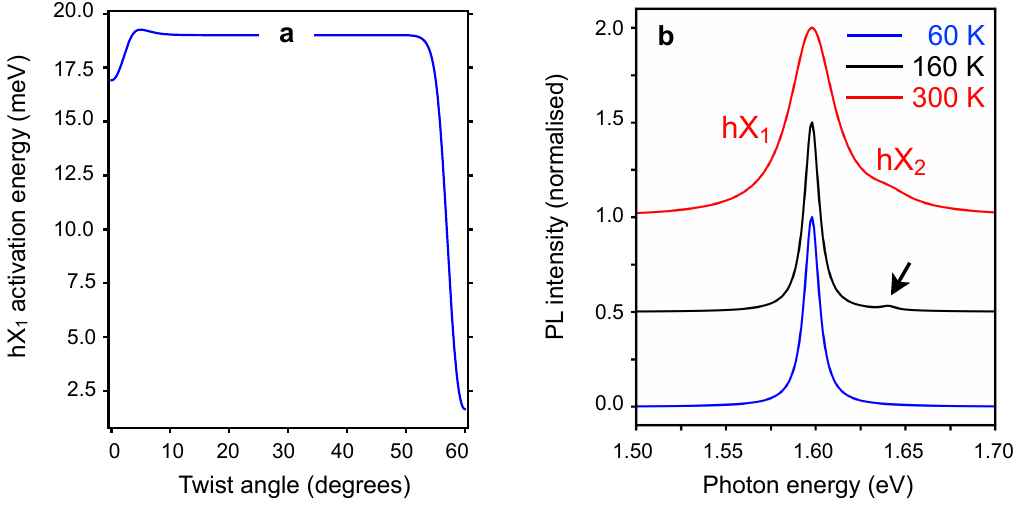}
	\caption{}
	\end{figure}
	
	%\newpage

\begin{figure}[hbt!]
	\centering
	\includegraphics[width=1\columnwidth]{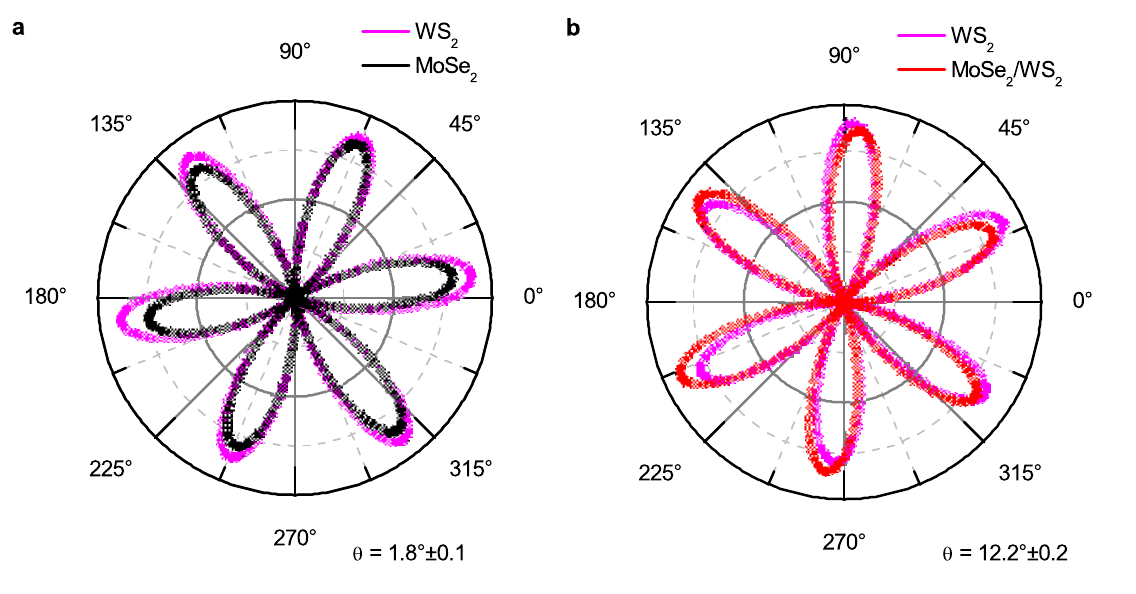}
	\caption{}
	\end{figure}

\end{document}